\documentclass[reprint,aps,prc,nofootinbib]{revtex4-2}

\usepackage{dcolumn}
\usepackage{dsfont}
\usepackage{mathtools}
\usepackage[english]{babel}
\usepackage{tikz}
\usepackage[compat=1.1.0]{tikz-feynman}

\usepackage{tensor}
\usepackage{graphicx} % Required for inserting images
\usepackage[utf8]{inputenc}
\usepackage{color}
\usepackage{slashed}
\usepackage{siunitx}
\usepackage{subcaption}
\usepackage{comment}
\usepackage{caption}

\begin{document}
\title{Towards a fluid-dynamic description of an entire heavy-ion collision: from the colliding nuclei to the quark-gluon plasma phase}
\author{Andreas~Kirchner}
\email{andreas.kirchner@duke.edu}
\affiliation{
Institut f\"{u}r Theoretische Physik, Ruprecht-Karls-Universit\"{a}t Heidelberg \\
Philosophenweg~16, D-69120 Heidelberg, Germany
}
\affiliation{Department of Physics, Duke University, Durham, NC 27708, USA}

\author{Federica Capellino}
\email{F.Capellino@gsi.de}
\affiliation{
GSI Helmoltzzentrum für Schwerionenforschung,\\
Planckstrasse 1,  64291 Darmstadt, Germany
}
\author{Eduardo Grossi}
\email[]{eduardo.grossi@unifi.it }
\affiliation{Dipartimento di Fisica, Universit\`a di Firenze and INFN Sezione di Firenze, via G. Sansone 1,
50019 Sesto Fiorentino, Italy}

\author{Stefan Floerchinger}
\email[]{stefan.floerchinger@uni-jena.de }
\affiliation{Theoretisch-Physikalisches Institut, Max-Wien-Platz 1, Friedrich-Schiller-Universit\"at Jena, 07743 Jena, Germany}

\newcommand*{\PFG}{P_{\mathrm{FG}}}
\newcommand*{\PBG}{P_{\mathrm{BG}}}

\newcommand*{\PWM}{P_{\mathrm{WM}}}

\newcommand*{\PHRG}{P_{\mathrm{HRG}}}
\newcommand*{\PLQCD}{P_{\mathrm{LQCD}}}
\newcommand{\ud}{\mathrm{d}}
\newcommand{\pib}{\pi_\mathrm{bulk}}

%\date{\today}

\begin{abstract}
The fluid-dynamical modeling of a nuclear collision at high energy usually starts shortly after the collision. A major source of uncertainty comes from the detailed modeling of the initial state. While the collision itself likely involves far-from-equilibrium dynamics, it is not excluded that a fluid theory of second order can reasonably well describe its soft features. Here we explore this possibility and discuss how the state before the collision can be described in that setup, what are the requirements from relativistic causality to the form of the equations of motion, how much entropy production can result from shear and bulk viscous dissipation during the initial longitudinal dynamics, and how one can thus obtain sensible initial conditions for the subsequent transverse expansion. While we do here only first steps, we outline a larger program. If the latter could be successfully completed it could lead to a dynamical description of heavy-ion collisions where the only uncertainty lies in the thermodynamic and transport properties of quantum chromodynamics.
\end{abstract}

\maketitle

\section{Introduction}

Ultra-relativistic heavy-ion collisions are a powerful tool to explore the QCD phase diagram~\cite{Guenther:2020jwe,Fukushima:2010bq} and its thermodynamic and transport properties. The extreme energy density achieved in such experiments likely allows for the transition of nuclear matter to a phase in which quarks and gluons are deconfined -- the quark-gluon plasma (QGP) \cite{Busza:2018rrf,ALICE:2010suc,PHENIX:2004vcz,STAR:2005gfr}. Relativistic viscous fluid dynamics based on the thermodynamic equation of state and transport properties of QCD is the established theory to describe large parts of the dynamical evolution of the fireball created by a high-energy nuclear collision~\cite{Baier:2006gy,Nijs:2020roc,Putschke:2019yrg,Romatschke:2017ejr}. Such a description characterizes the medium in terms of a few macroscopic fields (fluid velocity, temperature, chemical potentials, and dissipative currents). In the second-order fluid-dynamic formalism, the transport coefficients of the Navier-Stokes theory (shear and bulk viscosities, conductivities) are accompanied by corresponding relaxation times to ensure a causal evolution. The assumption of the fireball evolving like an expanding fluid successfully explained observables such as particle transverse momentum and azimuthal distributions~\cite{Nijs:2020roc,Schenke:2010nt,Floerchinger_2019}. In particular, radial and elliptic flow can be understood as a response to pressure gradients and signs for collective behavior.

The success of the fluid-dynamic description of many aspects of high-energy nuclear collisions~\cite{Giacalone:2023cet,JETSCAPE:2020mzn,Nijs:2020roc} motivates the search for the limits of its applicability~\cite{Carrington:2024utf,Inghirami:2022afu,Schlichting:2024uok,Bouras:2010hm,Molnar:2009pq,Niemi:2014wta,Ambrus:2022qya,Gallmeister:2018mcn,Noronha:2017oze}. One major challenge in this context is the description of the initial state. Typically, initial-state models aim to describe the system shortly after the overlap of the colliding nuclei. Phenomenologically, this is often done through geometrical assumptions as in the Monte-Carlo-Glauber model~\cite{Miller:2007ri}, where the shape of the overlap region of nucleonic sub-constituents at the time of impact is associated with the initial entropy density. The magnitude of the entropy density must be fixed by introducing a normalization constant that allows to reproduce the integrated yields of charged particles produced at the end of the fluid evolution. Initial conditions for other fluid fields like the fluid velocity, shear stress, or bulk viscous pressure are in practice simply postulated. These approaches, based on geometrical assumptions and a few parameters -- like the normalization factor for the entropy density or the initialization time $\tau_0$ -- of a fluid-dynamic description that are fitted to experimental data have been proven to be highly effective, e.g. in reproducing the correct hadron multiplicities as a function of centrality~\cite{Moreland:2014oya}. However, they do not provide a way to transition continuously from a description of the incoming nuclei to the dynamically evolving fireball.

In this paper, we propose to overcome the difficulty of formulating initial conditions for the fluid fields by extending the fluid-dynamic modeling of the heavy-ion collision to times before the collision. We argue that this might indeed be possible within second-order theories of relativistic fluid dynamics of the kind proposed by Israel and Stewart~\cite{Israel:1976tn}. It is conceivable that such theories remain (approximately) valid also relatively far from global thermal equilibrium and that they can describe the large amounts of entropy production needed for a transition from an initial state where entropy vanishes to a fireball of high temperature. Furthermore, expressing the entropy current of the fluid in terms of the state variables and dissipative currents, one can decompose the contribution of shear stress, bulk viscous pressure, and baryon diffusion current to the increase of entropy, relying on the assumption that this decomposition is valid even in the earliest stages of the heavy-ion collision. 

We will start with a fluid-dynamic description of the incoming nuclei in terms of their energy-momentum tensors and particle currents at vanishing temperature and finite baryon chemical potential. This allows us to explore the evolution of the fluid in the phase diagram and to naturally provide initial conditions for the fluid fields. It is important to emphasize here that the deviations from thermal equilibrium are substantial during the actual collision. The shear stress, bulk viscous pressure or baryon diffusion current can be so large that they cannot be considered as small corrections. Nevertheless, it is conceivable that a fluid dynamic description works reasonably well also during these violent stages, in which temperatures and chemical potentials are defined in the extended sense of a fluid approximation. Our approach then aims at describing the transition from the cold nuclear matter phase to the high-temperature and small chemical potential conditions that characterize the system produced in heavy-ion collisions at LHC and RHIC top-energies shortly after the collision.

Our proposal to treat the initial dynamics of a heavy-ion collision by means of fluid dynamics is not entirely new. There have been early attempts to describe the full collision with a single-fluid model (see e.g.~\cite{Clare:1986qj} for a review, and references therein). However, these models, being essentially based on ideal fluid dynamics, had major shortcomings that limited their applicability to center-of-mass collision energies of 100 - 1000 MeV per nucleon. Outside this regime, the approximation of local thermal equilibrium and zero mean free path could not be applied. As a solution to this, several so-called multi-fluid models have later been proposed, aiming to describe the full collision~\cite{Katscher:1993xs,Ivanov:2005yw,Cimerman:2023hjw,Karpenko:2023bok}. Here one introduces three fluids -- a projectile, a target and a fireball -- and their reciprocal interaction. Still, each fluid is otherwise assumed to be ideal, with a modified energy-momentum conservation law to account for some interaction between them. These so-called friction terms get parameterized phenomenologically.

In this paper, we attempt an alternative approach by treating the full collision system as one single viscous fluid, but in the sense of a relativistic second-order fluid theory. The role of ``friction terms'' is played by the standard dissipative terms like the shear and bulk viscosity and heat conductivity arising at first order in derivatives, as well as their second-order extensions like relaxation times. The crucial difference with respect to the previously proposed single-fluid models is given by the presence of dissipative corrections and additional degrees of freedom (the bulk viscous pressure, shear stress and diffusion current). This relaxes the assumption of local thermal equilibrium and allows for a mechanism of entropy production which is necessary to describe the initial condition of a heavy-ion collision. Furthermore, the assumption of finite mean free path is encoded in the presence of finite relaxation times, which naturally broadens the regime of applicability of the theory, ensuring at the same time a causal evolution of the fields.

Let us emphasize here that the theory we propose is not complete. We believe that it is of interest to make an attempt at a fluid-dynamic description of the soft QCD physics for an entire heavy-ion collision at high energies. We do several steps here towards such a description but more work is needed to determine whether it can work as a successful model of a heavy-ion collision.
While we believe that a dynamical description entirely based on the degrees of freedom of the energy-momentum tensor and other conserved currents (most prominently the one of baryon number) could work approximately, it is immediately clear that it cannot be exact. Hard QCD physics phenomena like highly energetic partons, or the jets they evolve into, for example, are not included.
It is also conceivable that the degrees of freedom of the energy-momentum tensor and conserved currents alone are not sufficient to fully capture the non-equilibrium state during the early collision phase. %sufficiently well with respect to soft dynamics.
If such an approach does work, however, one would gain a substantial amount of predictivity. Fluid properties like the equation of state, viscosities, conductivities, and relaxation times can at least in principle be determined from first principles using thermodynamics and response theory, as well as the fact that QCD is a renormalizable quantum field theory. The initial state before a collision is also fully fixed from equilibrium and simple kinematic considerations. In this sense, there are no free parameters. This attractive perspective serves as a motivation for the explorative work presented in this manuscript.

This paper is structured as follows: In Sec.~\ref{sec:NucleiAsStaticFluids} we outline a fluid-dynamic description of the incoming nuclei. On this basis we investigate in Sec.~\ref{sec:InteractionlessColl} a model for interactionless collisions corresponding formally to asymptotically high collision energies. The necessary thermodynamic equation of state is developed in Sec.~\ref{sec:EquationOfState}. Afterward, we develop a description of the collision system including interactions based on second-order viscous fluid dynamics in Sec.~\ref{sec:FluidDynModelling}. The equations of motion for the fluid fields are solved in Sec.~\ref{sec:Results} for a drop of nuclear matter undergoing an isotropic contraction and expansion. Finally, we draw some conclusions and formulate an outlook on possible extensions of our work in Sec.~\ref{sec:Outlook}.

\section{Nuclei as static fluids} \label{sec:NucleiAsStaticFluids}

\subsection{One nucleus}
Let us start by considering a single, isolated nucleus in vacuum. The baryon density can be modeled reasonably well by a Woods-Saxon distribution,
\begin{align}
    n(x,y,z)=\frac{n_0}{1+e^{(\sqrt{x^2+y^2+z^2}-R)/a}} ,\label{eq:InitalDensityWS}
\end{align}
where $R$ is the nuclear radius, %\footnote{This parametrization describes a spherically symmetric nucleus and can be extended to account for deformed nuclei by introducing an angle-dependent nuclear radius $R(\varphi,\theta)$}
$a$ is the skin parameter and $n_0$ is the nuclear saturation density. For nuclei that lack spherical symmetry, one can introduce further shape parameters.

From a thermodynamic point of view, standard nuclear matter finds itself on the first-order phase transition line at $T=0$ and $\mu = \mu_{\rm crit}$, with vanishing pressure, $p=0$. The baryon density of a homogeneous state jumps discontinuously from $n=0$ for $T=0$ and $\mu<\mu_{\rm crit}$ to the saturation density $n=n_0$ just above $\mu_{\rm crit}$. At the transition point itself average intermediate densities $0 \leq n \leq n_0$ can be realized in macroscopic volumes through phase separation. The energy $\epsilon$ and the number density $n$ are linked by $\epsilon=\mu_\text{crit} n$.  

Notice the importance of the first-order phase transition for the stability of the nucleus: If the nucleus were not at the phase transition and had a non-zero pressure, spatial gradients of the pressure would lead to dispersion. A deeper understanding of a nucleus can be attained by incorporating isospin-dependent terms into the thermodynamic framework, along with considering surface tension effects. From a fluid-dynamic perspective, surface tension is described in terms of higher-order derivatives. For the present exploratory work, we shall neglect such terms. Across the first-order phase transition one can then understand intermediate densities in terms of a superposition of phases within appropriate fractions of volume (see Sec.~\ref{sec:FluidDynModelling} for a more detailed discussion).

Given the energy density, baryon density, the (vanishing) pressure and the fluid velocity $u^\mu =(1,0,0,0)$ of a static fluid in its rest frame one can also compose the energy-momentum tensor and baryon current
\begin{align}
    T^{\mu\nu}=\epsilon u^\mu u^\nu + p\Delta^{\mu\nu}, \quad N^\mu = n u^\mu. \label{eq:IdTandN}
\end{align}
Here we choose the metric to have signature $(-,+,+,+)$ and define the projector orthogonal to the fluid velocity $\Delta^{\mu\nu}=g^{\mu\nu}+u^\mu u^\nu$. The static nucleus is in thermal equilibrium, hence no dissipative fields appear in Eq.~(\ref{eq:IdTandN}). The conservation laws are of course satisfied for the static configuration,
\begin{align}
    \nabla_\mu T^{\mu\nu}=0, \quad \nabla_\mu N^\mu =0. \label{eq:ConservationLaws}
\end{align}

This description can be readily extended to describe a nucleus moving with a constant velocity $\mathbf{v}$. By performing a Lorentz boost, the conservation equations become free-streaming equations,
\begin{align}
\label{eq:nucleus_eom}
    \partial_t \epsilon + v_i \partial_i \epsilon =0, \quad
    \partial_t n + v_i \partial_i n =0.
\end{align}
The solutions are given by
\begin{equation}
\label{eq:sol_moving}
    \epsilon(t,x) = \epsilon(0,x_i-v_it),\quad n(t,x)=n(0,x_i-v_it),
\end{equation}
where the initial baryon and energy density are transported with constant velocity. 
The fluid velocity $u^\mu$ has the components $u^0=\gamma$ and $u^i = \gamma v^i$, where $\gamma=(1-\mathbf{v}^2)^{-1/2}$.

In general, for a fluid in global thermal equilibrium, the second law of thermodynamics 
\begin{align}
\label{eq:second_law}
    \nabla_\mu S^\mu (x) \geq 0,
\end{align}
where $S^\mu$ is the entropy current, becomes an equality and the entropy current has to be stationary. This is the case when the global equilibrium conditions \cite{DeGroot:1980dk}
\begin{align}
    \nabla_\mu \beta_\nu + \nabla_\nu \beta_\mu = 0, \quad \partial_\nu \alpha =0 \label{eq:stabilityConditions}
\end{align}
for the ratios of the fluid velocity, $\beta^\mu = u^\mu /T$, and chemical potential to the temperature, $\alpha=\mu/T$, are satisfied. This is the case for nuclear matter at the liquid gas phase transition at vanishing temperature with constant fluid velocity. Dissipation can occur when the conditions in Eq.~(\ref{eq:stabilityConditions}) are violated.

\subsection{Two approaching nuclei}

Exploiting the description of the single boosted nucleus, it is possible to construct a collision system out of two nuclei moving towards each other. The energy-momentum tensor and number density current can be obtained by the superposition of two incoming nuclei with opposite constant momentum, initialized at $z = \pm z_0$ with $z_0 \gg R$ much larger than the nuclear radius. Here and in the following the coordinate system is defined such that the $z$-axis is along the beam pipe. The moment of full overlap of the two nuclei defines $t=0$ and the initial time $t_0$ is accordingly negative. In the following, the two nuclei boosted in the longitudinal (i.\ e.\ $z$-) direction with velocity $\pm v$ will be labeled with index $\rightarrow$ or $\leftarrow$. The energy-momentum tensor and particle density current are given by
\begin{align}
    N^\mu=N^\mu_\rightarrow + N^\mu_\leftarrow, \qquad T^{\mu\nu}=T^{\mu\nu}_\rightarrow+T^{\mu\nu}_\leftarrow, \label{eq:addingTandN}
\end{align}
and can, like any energy-momentum tensor with a time-like eigenvector, be further decomposed as
\begin{align}
\label{eq:Nmu}
    N^\mu &= n u^\mu +\nu^\mu, \\
    \label{eq:Tmunu}
    T^{\mu\nu} &= \epsilon u^\mu u^\nu +(p+\pib) \Delta^{\mu\nu} +\pi^{\mu\nu}.
\end{align}
We have chosen to define the fluid velocity $u^\mu$ in the so-called Landau frame~\cite{Landau1987Fluid} as a time-like eigenvector,
\begin{align}
    T^\mu_{\phantom{\mu}\nu} u^\nu = -\epsilon u^\mu,\label{eq:LandauMatchingEnergy}
\end{align}
with the eigenvalue given by the energy density $\epsilon$ of the composite system, and the fluid velocity normalized such that $u_\mu u^\mu =-1$. In Eq.~(\ref{eq:Nmu}), we introduced the diffusion current $\nu^\mu$, which is defined to be orthogonal to the fluid velocity ($u_\mu \nu^\mu =0$) and has therefore only three independent components. In Eq.~(\ref{eq:Tmunu}) we also introduced the bulk viscous pressure $\pib$ and the shear-stress tensor $\pi^{\mu\nu}$ orthogonal to the fluid velocity.

Note that the Landau frame matching expressed through Eq.~(\ref{eq:Nmu}) and Eq.~(\ref{eq:Tmunu}) becomes purely formal in the vacuum regions where the energy density vanishes. In these regions, the fluid velocity $u^\mu$ is not well defined, because any choice is an eigenvector of a vanishing energy-momentum tensor. These considerations also show that fluid velocity gradients in the vacuum regions cannot have any physical significance. An important question for our purpose is whether fluid dynamics can be organized such that it nevertheless remains consistent in these regions so that the colliding nuclei as well as the vacuum region around them can be formally seen as part of a fluid description. In the following, we discuss this question in the framework of dissipative relativistic fluid dynamics.

The simplest theory able to capture dissipative effects is the first-order relativistic Navier-Stokes theory, which is an extension of ideal fluid dynamics that provides constituent equations for the dissipative quantities appearing in the energy-momentum tensor and number density current. These constituent relations,
    $\pib = -\zeta \theta$,
    $\nu^\mu =-\kappa [ nT / (\epsilon+p)]^2\Delta^{\mu\nu}\partial_\nu \alpha$, and
    $\pi^{\mu\nu}= -2\eta \sigma^{\mu\nu}$, 
describe the dissipative fields as proportional to gradients of the fluid velocity $\theta=\nabla_\mu u^\mu$, $\sigma^{\mu\nu}=P^{\mu\nu\alpha}_{\phantom{\mu\nu\alpha}\beta}\nabla_\alpha u^\beta$ -- where $P^{\mu\nu\alpha}_{\phantom{\mu\nu\alpha}\beta}= [(1/2)\Delta^{\mu \alpha} \Delta^\nu_\beta +(1/2) \Delta^\mu_\beta \Delta^{\nu \alpha} -(1/3) \Delta^{\mu \nu}\delta^\alpha_\beta]$ -- and the gradient of the ratio of the chemical potential to the temperature $\alpha=\mu/T$. The proportionality coefficients, $\zeta$, $\kappa$, and $\eta$, are the bulk viscosity, heat conductivity and shear viscosity, respectively. All the gradients vanish in global thermal equilibrium, allowing to recover the ideal form of Eq.~(\ref{eq:Nmu}) and Eq.~(\ref{eq:Tmunu}). A second possibility for dissipative fields to vanish is in the limit of vanishing viscosities.

While the two nuclei are approaching each other and are at a large distance, they can be described as drops of nuclear matter separated by a vacuum region in the following way,
\begin{itemize}
    \item each nucleus has temperature $T=0$, chemical potential $\mu=\mu_{\rm crit}$, number density $n$ as in Eq.~(\ref{eq:InitalDensityWS}), energy density $\epsilon=\mu_{\rm crit}n$ and constant fluid velocity $v=\pm v_0$, in the region where $n>0$,
    \item the vacuum region, i.e. the space not occupied by the nuclei, has $T=0$, $\mu=\mu_{\rm crit}$, $n=0$, $\epsilon=0.$
\end{itemize}
The fluid velocity in $z$-direction is positive for one nucleus and negative for the other one; this implies that in between the nuclei, there is a region where the fluid velocity field displays non-vanishing gradients.
Within the Navier-Stokes theory, this implies that $\eta=\zeta=0$ in these regions so that the energy-momentum tensor in Eq.~(\ref{eq:Tmunu}) vanishes. The diffusion current can be written with $\epsilon=\mu n$ and $p=0$ as
\begin{align}
    \nu^\mu = -\kappa \left[ \frac{nT}{\mu} \Delta^{\mu\nu} \partial_\nu \mu + n \Delta^{\mu\nu} \partial_\nu T \right]
\end{align}
which vanishes for $\mu=\mu_\text{crit}$ and $T=0$ regardless of the specific choice of $\kappa$. This discussion can be generalized to second-order order fluid-dynamics (such as Israel-Stewart theory), as shown in Sec.~\ref{sec:Hydrodynamics}.

In summary, two approaching nuclei that do not yet overlap and the vacuum around them can be described in terms of fluid dynamics as long as $\eta=\zeta=0$ at $T=0$ and $n=0$. This picture breaks down as soon as the nuclei start to overlap since the fluid velocity has to transition between $+v$ and $-v$, with non-vanishing gradients, in a region where the densities are non-vanishing.

\section{Interactionless collision} 
%\section{Frictionless collision}
\label{sec:InteractionlessColl}

%\textcolor{brown}{[S.F. prefers ``interactionless collisions'' over ``frictionless collisions'' because frictionless can be either associated to the ideal fluid limit, which would be wrong, or in the multi-fluid models, which we do not introduce properly.]}

To better understand the description of a high-energy heavy-ion collision in terms of their energy-momentum tensor and conserved baryon number current, as decomposed in eqs.\ \eqref{eq:Nmu}, \eqref{eq:Tmunu}, we start with a simple exercise. We consider a hypothetical limit where the cross section between nuclei vanishes and aim to find the fluid fields according to the matching conditions in eqs.\ \eqref{eq:Nmu}, \eqref{eq:Tmunu}. This ``interactionless limit'' would correspond to vanishing friction terms in a multi-fluid model.

The conserved currents will here be modeled by a time-dependent linear superposition of the conserved currents (see Eq.~(\ref{eq:addingTandN})) with the nuclei centered at $z=\pm z_0 \mp v (t-t_0)$, i.e. $z=  \mp v t$ with $z_0 = -t_0>0$ in units where $c=1$. The resulting energy-momentum tensor and particle number density current can then be decomposed into the fluid fields using Landau frame matching. In this matching procedure, the fluid velocity $u^\mu$ and energy density $\epsilon$ are given by the time-like eigenvector and the corresponding eigenvalue of the energy-momentum tensor (Eq.~(\ref{eq:LandauMatchingEnergy})). The number density $n$ and diffusion current are obtained by solving Eq.~(\ref{eq:addingTandN}) for $n$ and the three independent components of $\nu^\mu$. The combination of pressure and bulk viscous pressure is found from the trace,
\begin{equation}
    p+ \pib = \frac{1}{3} \Delta^{\mu \nu} T_{\mu\nu}.
\end{equation}
To separate the bulk viscous pressure $\pib$ from the thermodynamic pressure $p$ one needs a thermodynamic equation of state in the form $p(\epsilon, n)$.
Finally, the shear stress tensor follows as
\begin{equation}
    \pi^{\mu\nu} = T^{\mu\nu} -\epsilon u^\mu u^\nu -[p+\pib] \Delta^{\mu\nu}.
\end{equation}

\begin{figure*}
        \centering
        \begin{subfigure}[b]{0.475\textwidth}
            \centering
            \includegraphics[width=1\textwidth]{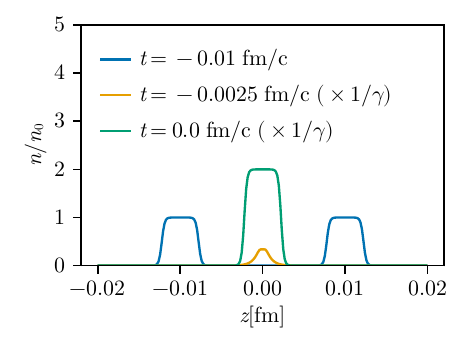}
            %\caption[Number density]%
            %{{\small Number density}}    
            \label{fig:LandauNumber}
        \end{subfigure}
        \hfill
        \begin{subfigure}[b]{0.475\textwidth}  
            \centering 
            \includegraphics[width=1\textwidth]{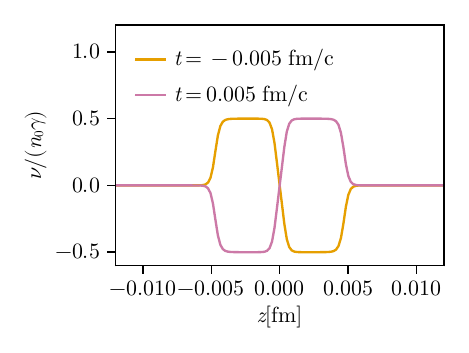}
            %\caption[]%
            %{{\small Diffusion current}}    
            \label{fig:LandauNu}
        \end{subfigure}
        \vskip\baselineskip
        \begin{subfigure}[b]{0.475\textwidth}   
            \centering 
            \includegraphics[width=1\textwidth]{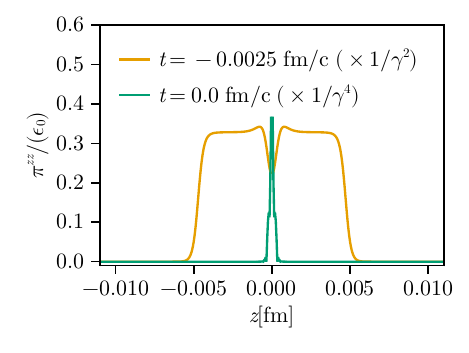}
            %\caption[]%
            %{{\small Shear stress}}    
            \label{fig:LandauPiZZ}
        \end{subfigure}
        \hfill
        \begin{subfigure}[b]{0.475\textwidth}   
            \centering 
            \includegraphics[width=1\textwidth]{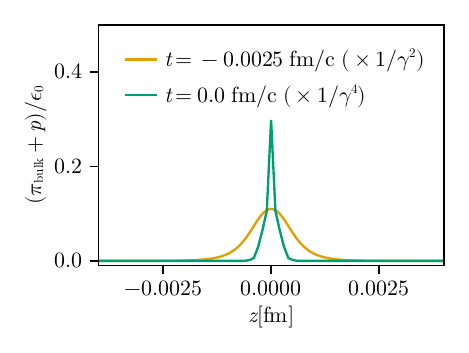}
            %\caption[]%
            %{{\small Bulk pressure}}    
            \label{fig:LandauPiB}
        \end{subfigure}
        \caption[]
        {\small Number density (top left), diffusion current (top right), shear stress (bottom left), and bulk pressure (bottom right) as a function of the spatial coordinate at different times of the collision in the interactionless limit, restricted to the longitudinal direction. The fluid fields are reconstructed from an effective energy-momentum tensor and number density current describing the two approaching nuclei, obtained as detailed in Eq.~(\ref{eq:addingTandN}). From these, the fluid fields describing the full collision system can be extracted by applying the decompositions Eq.~(\ref{eq:Nmu}) and Eq.~(\ref{eq:Tmunu}). The time dependence is modeled by carrying out the addition and matching procedure at $z=\pm z_0 \mp v (t-t_0)$, with the two nuclei initially (at $t_0$) being centered at $\pm z_0$ and approaching each other with $\mp v$. When the nuclei are far apart and can be treated as isolated the dissipative currents vanish. Additionally, the diffusion current vanishes by symmetry at full overlap. Note that the bulk pressure diverges at $z=0$ in this interactionless limit. Here and in the following the time is defined, such that $t=0$ is the moment of full overlap of the two nuclei.} 
        \label{fig:LongitudinalLandauMatching}
    \end{figure*}

Let us consider the example of a central ultra-relativistic Pb-Pb collision at $\sqrt{s_\text{NN}}=\SI{2.76}{\TeV}$. We restrict ourselves to its longitudinal dynamics in the $t$-$z$-plane at $x=y=0$. Here and in the following the time is defined, such that $t=0$ is the moment of full overlap of the two nuclei and the $z$-axis is defined to be along the beam pipe. The resulting number density $n$, bulk viscous pressure $\pib$ and independent components of the shear stress $\pi^{zz}$ and diffusion current $\nu$ are shown in Fig.~\ref{fig:LongitudinalLandauMatching}. Each field is plotted as a function of $z$ at different times. As long as the two nuclei are separated far enough to be considered independent, the non-equilibrium fields vanish. As soon as the two nuclei start to overlap ($t=\SI{-0.005}{\femto\meter /c}$), the number density increases and the two nuclei cannot be considered isolated anymore, resulting in non-vanishing stress fields. At full overlap, the bulk viscous pressure and shear stress reach their maximal values, while the diffusion current vanishes due to its odd parity transformation.

In the interactionless limit, the crossing of the two nuclei does not affect their structure. The dynamics of the nuclei at positive times ($t>0$) is analogous to the one of the incoming nuclei ($t<0$), with the two isolated nuclei moving apart from each other. The fluid velocity profile flips its sign after full overlap.

It is useful to further reflect on symmetries for a moment. For all symmetric collisions in the center-of-mass frame there is a discrete parity symmetry, $z\to -z$. For some fields, like the longitudinal component of the fluid velocity, that transforms like $v^z \to - v^z$ under such reflections, this implies a zero-crossing at $z=0$. We emphasize that this kind of apparent ``stopping'' is a consequence of the symmetries and the decomposition in eqs.\ \eqref{eq:Nmu}, \eqref{eq:Tmunu}, only, and does not tell anything about the validity of a fluid approximation. It also holds for the simple model of interactionless collisions.

For interactionless collisions there is a second discrete symmetry, namely with respect to time reflections, $t\to -t$ when time is measured relative to the moment of full overlap. This second symmetry holds only in very special situations where no entropy is produced. Another example besides the interactionless collisions would be a collision described by ideal fluid dynamics and we will encounter another case further below. However, for a realistic description of an inelastic heavy-ion collision, this symmetry will be broken.

From microscopic calculations, a thermodynamic equation of state is usually known in the grand canonical ensemble in the form of pressure as a function of temperature and chemical potential $p(T,\mu)$. The temperature and chemical potential can here be found by solving $\epsilon=\epsilon(T,\mu) = - p(T,\mu) + Ts+\mu n$, with $s(T,\mu)=\partial p(T, \mu)/ \partial T$ and $n=n(T,\mu) = \partial p(T,\mu)/\partial \mu$ for $T$ and $\mu$.
An equation of state covering a large part of the QCD phase diagram is needed for a consistent description of the full collision. We discuss some approximations for this in the next section.

\section{Thermodynamic equation of state in the temperature-chemical potential plane} \label{sec:EquationOfState}

\begin{figure}[h]
\includegraphics[width=0.5\textwidth]{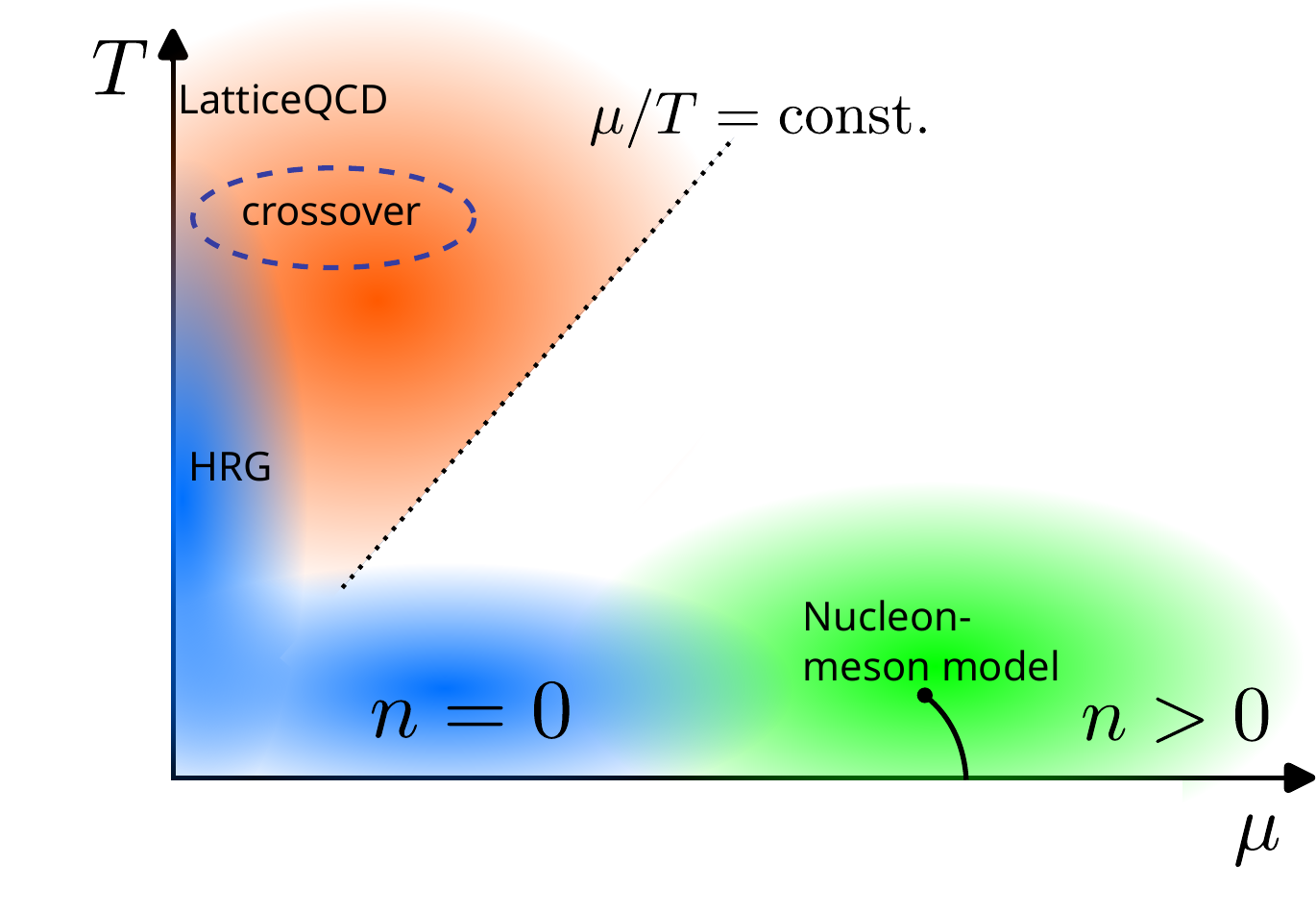}
\caption{Schematic QCD phase diagram. Different shaded areas correspond to the regime of validity 
of individual equation-of-state models.}
\label{fig:PhaseDiagramOverview}
\end{figure}

A fluid-dynamic description requires the specification of an equation of state, linking the densities appearing in the equations of motion, such as the energy density $\epsilon$ and baryon density $n$, to the pressure $p$ as well as to the state variables $T$ and $\mu$. The equation of state encodes microscopic physics, which is governed by QCD in case of a heavy-ion collision. There are several techniques to obtain information about an equation of state, including lattice gauge theory~\cite{HotQCD:2014kol,Borsanyi:2016ksw},  
effective theories or low-energy models~\cite{WALECKA1974491,Eser:2023oii}, or functional methods ~\cite{PAWLOWSKI2014113,Fu:2022gou}. All these methods usually provide an equation of state applicable only in a certain region of the QCD phase diagram and cannot cover it fully. Since our model aims at connecting the incoming nuclei of a heavy-ion collision (around $T=0$, $\mu=\mu_\text{crit}$) with the initialization point of traditional fluid-dynamic simulations at LHC or top-RHIC energies ($T\approx \SI{600}{\MeV}$, $\mu \approx 0$), an equation of state covering a large part of the phase diagram is needed. We will construct an approximate equation of state for the required region by combining results from lattice QCD, the hadron resonance gas (HRG), and a nucleon-meson model, which includes the first-order liquid-gas phase transition required for the stability of the incoming nuclei. A sketch of the QCD phase diagram displaying the areas of validity of each model is shown in Fig.~\ref{fig:PhaseDiagramOverview}. A more detailed discussion on the individual models can be found in App.~\ref{sec:appendixEoS}.

The equation of state shall be a smooth, differentiable function $p(T,\mu)$. (At the first order phase transition $n=\partial p/\partial \mu$ and $s=\partial p/\partial T$ are discontinuous.) We achieve this by connecting the lattice QCD (Sec.~\ref{sec:LQCD}) equation of state with the hadron resonance gas (Sec.~\ref{sec:HRG}) via a transfer function $f(T,\mu)$, inspired by Ref.~\cite{Monnai:2021kgu}, 
\begin{align}
    p(T,\mu)=&\frac{1}{2}(1-f(T,\mu))\PHRG(T,\mu)\nonumber\\
    &+\frac{1}{2}(1+f(T,\mu))\PLQCD(T,\mu), \label{eq:hybridEOS}
\end{align}
where $f$ is defined as
\begin{align}
    f(T,\mu)=\tanh\left(\frac{T-T_\text{trans}(\mu)}{\Delta T_\text{trans}} \right).
\end{align}
The transition temperature is chosen to be $T_\text{trans}(\mu) =\SI{0.1}{\GeV}+0.28\mu-\SI{0.2}{\GeV^{-1}}\mu^2 $ together with $\Delta T_\text{trans} = 0.1 T_\text{trans}(\mu=0)$. Around the phase transition line, the equation of state is switched from Eq.~(\ref{eq:hybridEOS}) to the nucleon-meson model (Sec.~\ref{sec:Walecka}). To overcome the sign problem, the lattice QCD equation of state uses a Taylor expansion in $\mu/T$ around $\mu=0$ to account for the regions with finite baryon density. The expansion can be used for $\mu/T \lesssim 3.5$. For low temperatures, the equilibrium part matches the hadron resonance gas and transitions to deconfined quarks and gluons at $T_c\approx \SI{155}{\MeV}$. 

\begin{figure*}
        \centering
        \begin{subfigure}[b]{0.475\textwidth}
            \centering
            \includegraphics[]{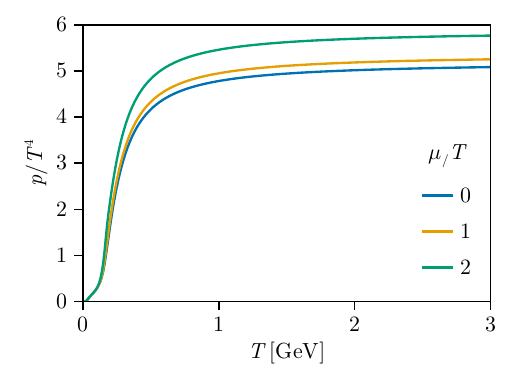}
            %\caption[Number density]%
            %{{\small Pressure in units of $T^4$ as a function of the temperature $T$ for different values of the chemical potential to temperature ratio, in the small $\mu /T$ domain.}}    
            %\label{fig:mean and std of net14}
        \end{subfigure}
        \hfill
        \begin{subfigure}[b]{0.475\textwidth}  
            \centering 
            \includegraphics[]{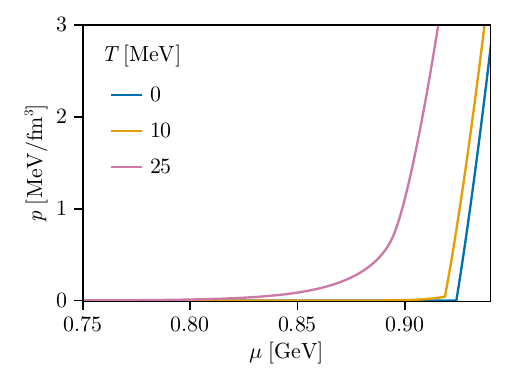}
            %\caption[]%
            %{{\small Pressure as a function of the chemical potential $\mu$ for different values of temperature $T$, in the large $\mu /T$ domain.\phantom{One more line needed for formatting}}}    
            %\label{fig:mean and std of net24}
        \end{subfigure}
        \caption[]
        {\small Composite equation of state for the low (left panel) and high (right panel) $\mu /T$ regions of the QCD phase diagram. At high temperatures, the pressure in units of $T^4$ is flat, as expected from the free quark model. At low temperatures, the pressure displays a sudden increase, due to the first-order gas-liquid phase transition.} 
        \label{fig:pressureComp}
\end{figure*}

In Fig.~\ref{fig:pressureComp} we show the results of the interpolation for $p/T^4$ at small $\mu/T$ values as a function of temperature (left panel) and the pressure as a function of the chemical potential for large values of $\mu/T$ (right panel).

\begin{figure*}
        \centering
        \begin{subfigure}[b]{0.475\textwidth}
            \centering
            \includegraphics[width=1\textwidth]{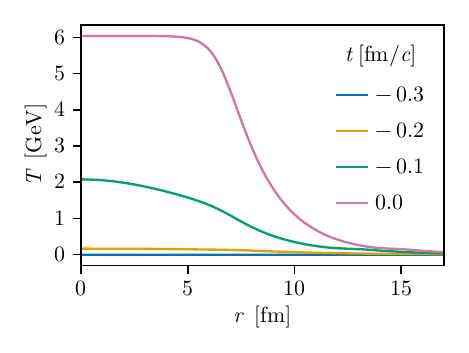}
            %\caption[Number density]%
            %{{\small Temperature}}    
            \label{fig:mean and std of net14}
        \end{subfigure}
        \hfill
        \begin{subfigure}[b]{0.475\textwidth}  
            \centering 
            \includegraphics[width=1\textwidth]{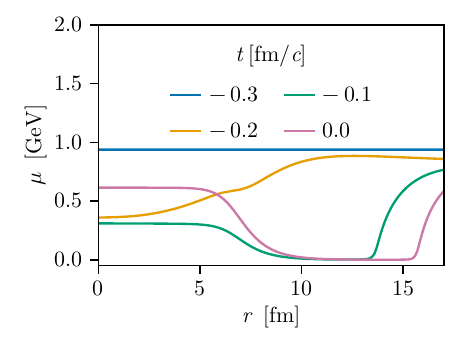}
            %\caption[]%
            %{{\small Chemical potential}}    
            \label{fig:mean and std of net24}
        \end{subfigure}
        \caption[]
        {\small Temperature (left panel) and chemical potential (right panel) as function of the transverse radius, evaluated at different times, with $t=0$ corresponding to the full overlap of the two nuclei. In the central region, i.e. for small $r$, the temperature and chemical potential are almost constant at all times.} 
    \label{fig:LongitudinalLandauMatchingTandMu}
\end{figure*}

With this equation of state, we can study the evolution of the temperature and chemical potential in the interactionless collision limit. The results are shown in Fig.~\ref{fig:LongitudinalLandauMatchingTandMu}. Initially, the two nuclei are separated enough to be considered isolated, sitting at $T=0$ and $\mu=\mu_\text{crit}$. When beginning to overlap, the temperature obtained from the matching in Eq.~(\ref{eq:Nmu}), Eq.~(\ref{eq:Tmunu}), and the equation of state increases due to the increased energy density in the overlap region. Concurrently, this leads to a decrease in the chemical potential. Close to full overlap, the dissipative currents either vanish by symmetry ($\nu=\nu^z$) or become delta-like peaks ($\pib$, $\pi^{zz}$), again yielding an increase of the chemical potential. 

As a function or radius $r=\sqrt{x^2+y^2}$, evaluated in a co-moving slice of the nucleus at $z=z_0+v(t-t_0)$, we find only small variations in the temperature and chemical potential in the center of the nucleus but then a decay towards the vacuum for larger radii, as expected.  

While the Landau frame matching consideration for interactionless collisions gives a first idea about the expected fluid fields during the collision, it is of course not realistic. A better picture can be obtained by employing viscous fluid dynamics in a second-order formulation, yielding a non-symmetric time evolution. The equations for the dissipative case will be developed in the next section.

\section{Fluid dynamic model for the collision itself} \label{sec:FluidDynModelling}

The basis of any fluid-dynamic description is given by conservation laws, which in turn are consequences of symmetries. In the case of a heavy-ion collision, the conservation of energy and momentum, as well as baryon number, are most important. This can be expressed in terms of the conservation laws in Eq.~(\ref{eq:ConservationLaws}). 
Together with a decomposition as in Eq.~(\ref{eq:Nmu}) and Eq.~(\ref{eq:Tmunu}) this leads to the following evolution equations for energy density, fluid velocity, and baryon number density
\begin{align}
    u^\mu \partial_\mu \epsilon + (\epsilon + p +\pib) \nabla_\mu u^\mu +\pi^\mu_{\phantom{\mu}\nu} \nabla_\mu u^\nu &=0,\label{eq:energy1}\\
    (\epsilon +p +\pib) u^\mu \nabla_\mu u^\nu \nonumber\\ + \Delta^{\mu\nu} \partial_\mu (p+\pib) + \Delta^\nu_{\phantom{\nu}\alpha} \nabla_\mu \pi^{\mu\alpha}&=0, \label{eq:energy2}\\
    u^\mu \partial_\mu n + n \nabla_\mu u^\mu +\nabla_\mu \nu^\mu &=0.\label{eq:number}
\end{align}
Alternatively, an equivalent formulation describing the time evolution of the temperature and chemical potential can be found with a change of variables using the thermodynamic differentials of the energy and number density,
\begin{align}
    \ud \epsilon &= \left[T \frac{\partial^2 p}{\partial T^2} + \mu \frac{\partial^2 p}{\partial T \partial \mu}\right] \ud T + \left[ T \frac{\partial^2 p}{\partial T \partial \mu}+ \mu \frac{\partial^2 p}{\partial \mu^2} \right] \ud \mu, \\
    \ud n &= \frac{\partial^2 p}{\partial T\partial\mu} \ud T + \frac{\partial^2 p}{\partial\mu^2} \ud \mu .
\end{align}

Evolving $T$ and $\mu$ instead of $\epsilon$ and $n$ has technical advantages because thermodynamic and transport properties are usually available from calculations in the grand canonical ensemble, i.\ e.\ as a function of $T$ and $\mu$.

Special considerations are needed when the evolution crosses a first-order phase transition. During the transition, the temperature and chemical potential of the two phases are identical, while the relative proportion of volume in each phase changes. Therefore, an additional parameter is required for the description of the evolution on the phase transition line.

In principle, the dynamics of the transition is governed by bubble nucleation in the metastable region or spinodal decomposition. For our present purpose, we are following a simpler description for macroscopic length scales and slow enough dynamics (in the spirit of fluid dynamics) and assume that the phase transition gets instantaneously triggered when the phase transition line is reached. The additional parameter describing the phase transition is the ratio of the volume of one of the phases to the total volume, $r\equiv V''/V$. A practical way to implement such a description is via the Massieu potential density $w$ and its differential,

\begin{align}
    w\left(\beta, \alpha\right) &= \beta p,\\
    \mathrm{d} w &= -\epsilon \mathrm{d} \beta + n \mathrm{d} \alpha,
\end{align}
where $\beta = 1/T$ and $\alpha = \mu/T$. For mechanical stability, the values of pressure p, and therefore the values of the Massieu potential $w$, agree in the two coexisting phases. Derivatives with respect to $\beta$ and $\alpha$ (energy and particle density) do not need to agree, however. The Massieu potential for a volume $V$ with subvolume $V'=(1-r)V$ in phase I and subvolume $V''=rV$ in phase II is given by
\begin{align}
     w(\gamma) = [1-r] \, w^\prime(\gamma) + r\,  w^{\prime\prime}(\gamma),
\end{align}
where $w^\prime$ and $w^{\prime\prime}$ are the Massieu potentials associated with phases I and II, respectively. Here we express the Massieu potential in terms of coordinates $\gamma^m =( -\beta,\alpha)$, which allows a geometric thermodynamic description. If $r=0$ and $w=w^\prime(\gamma)$ the system is in phase I. Where $r=1$ and  $w=w^{\prime\prime}(\gamma)$ the system lies fully in phase II. 

Expressing the conserved charge densities as $c_m=(\epsilon,n)$ allows to write at the phase transition
\begin{align}
    c_m(\gamma) = [1-r] \, c^\prime_m(\gamma) + r \, c_m^{\prime\prime}(\gamma),
\end{align}
with the corresponding differential,
\begin{equation}
\begin{split}
  d c_m =  \left[ (1-r) \, G^\prime_{mn}(\gamma) + r \, G^{\prime\prime}_{mn}(\gamma) \right] &d\gamma^n\\
  + [c_m^{\prime\prime}(\gamma) - c_m^\prime(\gamma)] &dr.
  \end{split}
  \label{eq:conservedDifferentialPhaseTransition}
\end{equation}
The quantities $G_{mn}^\prime$ and $G_{mn}^{\prime\prime}$ are known as thermal Fisher metrics in phase I and phase II, respectively. They are defined as 
\begin{equation}
  G_{mn}(\gamma) = \frac{d^2w(\gamma)}{d\gamma^m d\gamma^n} = \frac{d c_n}{d \gamma_m}.
  \label{eq:defFisherMetric}
\end{equation}
The thermal Fisher metric also allows to express the conservation equation of the conserved charge densities $c_m$ as
\begin{equation}
    u^\mu \partial_\mu c_m+f_m = 0,\label{eq:evConservedCharge}
\end{equation}
where the $f_m$ involves all the fluid fields and their derivatives. In terms of the conjugated, intensive variables $\gamma^m$, the conservation laws imply %read as

\begin{equation}
  u^\mu \partial_\mu \gamma^n + G^{nm}(\gamma, r) \left[ f_m + [c_m^{\prime\prime}(\gamma)-c_m^\prime(\gamma)] u^\mu \partial_\mu r \right] = 0.
  \label{eq:evgamman}
\end{equation}
Here $G^{nm}(\gamma, r)$ is the inverse of the linear combination of thermal Fisher metrics appearing in Eq.~(\ref{eq:conservedDifferentialPhaseTransition}).

To obtain an evolution equation for the volume ratio parameter $r$, we exploit that the thermodynamic variables remain constrained to the hypersurface of the phase transition, as long as the transition is not complete. Therefore, one can write down the condition
\begin{equation}
    n_m(\gamma) \ud \gamma^m =0, \quad \text{for}\quad 0 < r < 1,
\end{equation}
where $n_m(\gamma)$ is the normal vector of the transition hypersurface. The resulting evolution equation for the volume ratio parameter is then given by,
\begin{equation}
  u^\mu \partial_\mu r = - \frac{n_p(\gamma) G^{pq}(\gamma, r) f_q}{n_r(\gamma) G^{rs}(\gamma, r) [c_s^{\prime\prime}(\gamma) - c_s^\prime(\gamma)]}. \label{eq:EvolutionVolumeRatioParam}
\end{equation}
By inserting Eq.~(\ref{eq:EvolutionVolumeRatioParam}) into equation \eqref{eq:evgamman} for the coordinates $\gamma^n$, one finds

\begin{widetext}
 \begin{equation}
  u^\mu \partial_\mu \gamma^n + G^{nm}(\gamma, r) \left[ f_m - [c_m^{\prime\prime}(\gamma) - c_m^\prime(\gamma)] \frac{n_p(\gamma) G^{pq}(\gamma, r) f_q}{n_l(\gamma) G^{ls}(\gamma, r) [c_s^{\prime\prime}(\gamma) - c_s^\prime(\gamma)]}  \right]=0. \label{eq:ModifiedEvolutionCoords}
\end{equation}
\end{widetext}

This closes the set of thermodynamic evolution equations across the phase transition. The presence of additional conserved charges and their corresponding chemical potentials can be accounted for easily with this formalism. To fully determine the evolution of the fluid fields, supplemental equations for the dissipative currents need to be provided.

\subsection{Second-order fluid dynamics}
\label{sec:Hydrodynamics}

The equations of motion for the thermodynamic variables and the fluid velocity Eq.~(\ref{eq:energy1}), Eq.~(\ref{eq:energy2}) and Eq.~(\ref{eq:number}), which implement the conservation laws Eq.~(\ref{eq:ConservationLaws}), can only be solved when additional equations are provided for the shear stress $\pi^{\mu\nu}$, the bulk viscous pressure $\pib$ and the diffusion current $\nu^\mu$. 

Relativistic causality must be respected, which is the case when the evolution equations are hyperbolic with characteristic velocities below the speed of light. For the actual collision, one can expect relatively strong gradients and possibly shock-like behavior. This poses additional requirements for any viable form of fluid evolution equations. We would expect that these requirements are so strong that many proposed fluid theories fail, but it is conceivable that some succeed and, in turn, lead at least to a qualitatively correct and approximate quantitative description.

We will now discuss the set of equations proposed by Israel and Stewart and argue that they might have a prospect to lead to a complete (albeit approximate) description of a heavy-ion collision from before the collision to particle freeze-out.

First, recall that the diffusion current, the bulk viscous pressure, and the shear stress tensor vanish in equilibrium and parameterize therefore deviations out-of-equilibrium. Their dynamical evolution is not constrained by the conservation laws Eq.~(\ref{eq:ConservationLaws}) but must be supplemented by other considerations. A reasonable assumption able to constrain equations of motion for these fields, originally proposed by Israel and Stewart \cite{14-moment}, is based on a local version of the second law of thermodynamics as in Eq.~\eqref{eq:second_law}.
%\begin{equation}
%\label{eq:second_law}
%    \nabla_{\mu} S^\mu \geq 0.
%\end{equation}
The entropy current $S^\mu(x)$ is here assumed to be an algebraic and local functional of the fluid fields, i.e. the degrees of freedom of the energy-momentum tensor and baryon number current as decomposed in Eq.~(\ref{eq:Nmu}) and Eq.~(\ref{eq:Tmunu}). Two assumptions are needed to find such a current: the first is that it reduces to the entropy density in equilibrium, where the dissipative currents are set to zero. The second is that the dependence on the non-equilibrium variables is algebraic, and a perturbation in some sense. The latter is a working hypothesis needed to obtain a system of equations with just a few parameters (the second-order transport coefficients). 
Within these two hypotheses, and by exploiting the equations of motion and the first law of thermodynamics, %the divergence of the  entropy current reads,
the entropy current reads,
\begin{equation}
\begin{split}
S^\mu= s u^\mu - \frac{\mu}{T} \nu^{\mu} -Q^{\mu},
\end{split}
\end{equation}

where the entropy density at equilibrium is given by
\begin{equation}
s = \frac{1}{T}(\epsilon+P-\mu n) \,,
\end{equation}
and $Q^{\mu}$ is a four-vector that vanishes in equilibrium and represents the out-of-equilibrium contribution to the entropy current. 
The actual form of the out-of-equilibrium entropy current must be chosen such that its divergence is always positive when the equations of motion are obeyed. The simplest assumption, almost quadratic in the non-equilibrium fields, is used to fix the equations of motion that the dissipative currents must satisfy. 
The non-equilibrium entropy in the Landau frame can be taken as 
\begin{align}
    T Q^{\mu} = &\frac{u^\mu}{2}\left[\frac{\tau_\text{bulk}}{\zeta} \pib^2 +\frac{\tau_\text{heat}(\epsilon+p)^2}{\kappa n^2T^2} \nu^2 +\frac{\tau_\text{shear}}{2 \eta} \pi^2\right] \nonumber\\
    &-\alpha_0 \pib \nu^{\mu} -\alpha_1 \pi^{\mu\alpha}\nu_\alpha,
\end{align}
where we introduced the bulk viscosity $\zeta$, the heat conductivity $\kappa$, the shear viscosity $\eta$, and the transport coefficients $\alpha_0$ and $\alpha_1$, which may generically depend on the temperature and the chemical potential. Notice that, since the square of any dissipative field contributes positively to $Q^\mu$, the coefficients $\zeta$, $\kappa$, and $\eta$ must be positive, such that the entropy is maximized in equilibrium where such contributions vanish. The sign of $\alpha_0$ and $\alpha_1$, on the contrary, cannot be established \textit{a priori}. The divergence of the entropy current under this assumption for $Q^\mu$ reads,

    \begin{equation}
    \label{eq:entropy}
    \nabla_{\mu}S^{\mu} = \frac{\pib^2}{\zeta T}+ 
    \frac{\pi^{\mu\nu}\pi_{\mu\nu}}{2\eta T}
    +\frac{(\epsilon+p)^2 \nu^\mu\nu_\mu}{\kappa (nT)^2 }.
    \end{equation}
By imposing Eq.~(\ref{eq:second_law}) and the aforementioned form for the dissipative contribution, one finds the following equations of motion for the bulk viscous pressure $\pib$,
\begin{align}
\label{eq:IS}
\begin{split}
     \tau_\mathrm{bulk} u^\mu \partial_\mu \pib +\pib = -\zeta \bigg[\theta+\alpha_0\nabla_\mu \nu^\mu  & \\  +\frac{1}{2}T\nabla_\mu \left(\frac{\tau_\mathrm{bulk} u^\mu}{\zeta T}\right )\pib   + \frac12 T \nu^\mu \partial_\mu \left(\frac{\alpha_0}{T} \right) \bigg]&,
\end{split}
\end{align}
for the diffusion current $\nu^\rho$ 
\begin{align}
\begin{split}
     \tau_\mathrm{heat} \Delta^{\rho}_{\phantom{\rho}\sigma}u^\mu \nabla_\mu \nu^\sigma +\nu^\rho =  -\kappa\left[\frac{nT}{\epsilon+p}\right]^2 & \\ \times \bigg[  \Delta^{\rho\sigma}\partial_\sigma \left(\frac{\mu}{T}\right)
     + \frac{1}{2}  \nu^\rho \nabla_\mu \left(\frac{(\epsilon +p)^2\tau_\mathrm{heat}}{\kappa n^2 T^2} u^\mu \right) & \\ + \frac{\alpha_0}{T}\Delta^{\rho\mu}\partial_\mu \pi_\text{bulk}  +\frac{\alpha_1}{T}\Delta^{\rho}_{\phantom{\rho}\sigma}\nabla_\mu\pi^{\mu\sigma} & \\ +\frac12   \Delta^{\rho\mu} \partial_\mu \left(\frac{\alpha_0}{T} \right)\pi_\text{bulk}  +\frac12  \pi^{\rho\mu} \partial_\mu\left(\frac{\alpha_1}{T} \right)\bigg]&,\label{eq:IS2}
\end{split}
\end{align}
and for the shear stress $\pi^{\mu\nu}$
\begin{align}
\begin{split}
     \tau_\mathrm{shear} P^{\mu\nu}_{\phantom{\mu\nu}\alpha\beta} u^\rho \nabla_\rho \pi^{\alpha\beta} + \pi^{\mu\nu}= & \\ -\eta \bigg[2\sigma^{\mu\nu}+ 2\alpha_1 P^{\mu\nu\alpha}_{\phantom{\mu\nu\alpha}\beta}\nabla_\alpha\nu^\beta   & \\ +  T \pi^{\mu\nu} \nabla_\rho \left(\frac{\tau_\mathrm{shear}}{2\eta T} u^\rho\right)  +TP^{\mu\nu\rho}_{\phantom{\mu\nu\rho}\sigma}\partial_\rho \left(\frac{\alpha_1}{T} \right)\nu^\sigma \bigg]&.\label{eq:IS3}
     \end{split}
\end{align}
The symbol $P$ is the projector into the symmetric transverse traceless subspace to the four-velocity, 
\begin{equation}
    P^{\alpha \beta}_{\phantom{\alpha\beta}\lambda\mu} = \frac12\Delta^\alpha_{\phantom{\alpha}\lambda}\Delta^\beta_{\phantom{\beta}\mu} + \frac12 \Delta^\beta_{\phantom{\beta}\lambda}\Delta^\alpha_{\phantom{\alpha}\mu} -\frac13 \Delta^{\alpha\beta} \Delta_{\lambda\mu}, 
\end{equation}
and the shear tensor $\sigma^{\mu\nu}$ is defined as 
\begin{equation}
    \sigma^{\lambda\mu }= P^{\lambda\mu\alpha}_{\phantom{\lambda\mu\alpha}\beta} \nabla_{\alpha}u^\beta. 
\end{equation}
Eq.~(\ref{eq:IS}), Eq.~(\ref{eq:IS2}), and Eq.~(\ref{eq:IS3}) are often referred to as Israel-Stewart (IS) equations and are relaxation-type equations. They reduce to the Navier-Stokes' constitutive equations formally for large evolution times or small relaxation times ($t\gg \tau_\mathrm{bulk}, \tau_\mathrm{shear}, \tau_\mathrm{heat}$). The presence of relaxation times ensures the causal behavior of the fluid fields, assuming that they are sufficiently large. Typically the relaxation times become very large in dilute regions where collisions between particles are rare.

To employ the evolution equations Eq.~(\ref{eq:IS}), Eq.~(\ref{eq:IS2}), and Eq.~(\ref{eq:IS3}) for a consistent description of freely propagating nuclei as nuclear matter droplets, we have to ensure that the dissipative currents are consistently vanishing in the vacuum region. Within the Israel-Stewart formalism, vanishing viscosities $\eta=\zeta=0$ at $T=0$ and $n=0$ is not the only solution. For diverging relaxation times $\tau_\text{bulk}, \tau_\text{shear} \to \infty$ with $\eta/\tau_\text{shear}=\zeta/\tau_\text{bulk}=0$ in this limit, the shear stress and bulk viscous pressure become time-independent. Accordingly, with the additional assumption that they vanish at very early times, this provides an alternative possibility.

Furthermore, the limiting case of infinite relaxation times shows that a solution to the equation of motion corresponding to two nuclei flying through each other -- i.e. complete \textit{transparency} -- is achievable within second-order dissipative theories. The issue of how to achieve transparency was, historically, one of the reasons that set the single-fluid description aside, in favor of multi-fluid models~\cite{Clare:1986qj}. Below we argue that a second-order theory can naturally overcome this problem. The realistic case of finite relaxation time, corresponding to a non-zero but finite mean free time, will allow for a modification of the dissipative currents, resulting in a non-complete transparency of the two colliding nuclei.

As a further remark, we emphasize that Eq.~(\ref{eq:entropy}), supplemented with Israel-Stewart equations, allows us to estimate the contribution to the entropy production coming from the different sources of dissipation (shear and bulk viscous dissipation as well as heat conduction or baryon diffusion). This is particularly interesting in the context of heavy-ion collisions, where the initial entropy density on a Cauchy surface is usually fixed by hand shortly after the collision to reproduce the charged particle yields measured by the experiment. In contrast, if a fluid description can be employed at all times, all entropy must be produced according to Eq.~(\ref{eq:entropy}). For given fluid properties, one could even see this as an important test for the validity of the fluid description.

\subsection{Longitudinal collision dynamics} \label{sec:LongitudinaCollision}

In the early stages of the collision, it is expected that the dynamics is dominated by gradients in the longitudinal directions. This is a consequence of the strong longitudinal Lorentz contraction along the beam axis, leading to larger gradients of the number and energy density. This assumption is supported by the interactionless limit (Fig.~\ref{fig:LongitudinalLandauMatchingTandMu}), showing vanishing transverse gradients of the temperature and chemical potential in the central overlap region. A reasonable assumption is therefore to only account for gradients in the longitudinal direction. This is equivalent to assuming the collision of two nuclei with an infinite extent in the transverse plane, resulting in the invariance under translations in addition to rotations around the beam axis. 

These symmetries reduce the number of independent components of the fluid fields, resulting in
\begin{align}
    u^\mu &= (\gamma,0,0,\gamma v),\\
    \nu^\mu &=(v \nu,0,0,\nu),
\end{align}
where the normalization of the fluid velocity, $u_\mu u^\mu =-1$, and its orthogonality to the diffusion current, $u_\mu \nu^\mu=0$, were used. Similarly, the shear-stress tensor can be parametrized as
\begin{align}
    \pi^{\mu\nu}= \begin{pmatrix}
v^2 & 0 & 0 & v\\
0 & -\frac{1}{2\gamma^2}  & 0 & 0 \\
0 & 0 & -\frac{1}{2\gamma^2}   & 0 \\
v & 0 & 0 & 1
\end{pmatrix}\pi^{zz},
\end{align}
after using its orthogonality to the fluid velocity and tracelessness. In this parametrization, the six independent fluid fields are $\Phi=(T,\mu,v,\pib,\pi^{zz},\nu)$. The corresponding equations of motion are first-order partial differential equations of the type
\begin{align}
\label{eq:quasilinear}
    A_{ij}(\Phi) \partial_t \Phi_j +B_{ij}(\Phi) \partial_z \Phi_j +S_i(\Phi)=0,
\end{align}
where $A_{ij}$, $B_{ij}$, and $S_i$ are the coefficient matrices and the source term, respectively, and can all depend on the fluid fields.

The fluid velocity $v$ is antisymmetric with respect to $z\to -z$ and therefore vanishes at $z=0$. One expects a Bjorken-type flow profile, i.\ e.\ $v = z/t$ around $z=0$, to emerge at some time $t$ after the collision. Here that should arise as a result of the solution to the equations of motion instead of being an assumption.

Determining solutions to partial differential equations of the type of Eq.~\eqref{eq:quasilinear} for highly energetic collisions is challenging. The initial state can be seen as posing a kind of Riemann problem where the density is approximately constant, the temperature is vanishing, and the velocity is piecewise constant. Classical solutions as continuous and differentiable functions, can likely not be expected. 

The solution to the Riemann problem for the Euler equation (ideal fluid dynamics) requires additional information besides the equations of motion. Weak solutions are \textit{a priori} not unique but the physical solution is singled out by the Rankine-Hugoniot condition that ensures that entropy is non-decreasing.

The ideal fluid equations have the form of conservation laws as in Eq.~\eqref{eq:ConservationLaws}, which is the decisive feature that allows us to identify weak solutions and to find them numerically. 

Therefore, if one neglects the shear, bulk, and heat conductive dissipation and uses ideal fluid dynamics, one likely can find appropriate weak solutions to the Riemann problem. While the equations of motion \eqref{eq:quasilinear} are in general not of conservative type, they still retain hyperbolic form with associated non-decreasing entropy, for a large class of states \cite{Floerchinger:2017cii}.

The description of a longitudinally expanding system with ideal fluid dynamics was already undertaken by Landau \cite{Landau:1953gs}, see Ref.~\cite{florkowski2010phenomenology} for a review. The initial condition in this case is given by a thick disk of matter, to resemble the result of an initial compression stage. This model allows to find Bjorken flow as an approximate late-time solution for the equations of motion. Landau's description is limited to a simple equation of state and ideal fluids. In this work, we are interested in the evolution starting from normal nuclear matter and therefore also follow the baryon currents. Furthermore, dissipation seems crucial, in order for the system to transition from a low $T$/high $\mu$ initial condition to a high $T$/low $\mu$ fluid regime dynamically, and also second-order relaxation times play an important role.
%For the problem of colliding nuclei and within ideal fluid dynamics this was already undertaken by Landau \cite{Landau:1953gs}. 

In the original formulation of Israel and Stewart \cite{14-moment}, as well as later refinements, \cite{Denicol:2012cn}, the equations are hyperbolic but can not be put in conservative form. Intuitively, the necessary condition for that would be that the matrices $A$ and $B$ in \eqref{eq:quasilinear} can be integrated inside the time and space derivatives. For a single equation, 
this is always possible, but for a system of equations, the matrices need to satisfy certain integrability conditions. 

To understand this we consider the divergence-type equations 
\begin{equation}
    \partial_t U_i + \partial_k F_i^k + S_i = 0 ,
\label{eq:conservativeForm}
\end{equation}
where we assume that $U$, $F$ and $S$ depend on the variables $\Phi_j$. Taking a  derivative we have 
\begin{equation}
    \frac{\partial U_i}{\partial \Phi_j} \partial_t \Phi_j+  \frac{\partial F^k_i}{\partial \Phi_j} \partial_k \Phi_j +S_i=0 \;.
\end{equation}
By comparing to Eq.~\eqref{eq:quasilinear}, we have 
\begin{equation}
    A_{ij} = \frac{\partial U_{i}}{\partial \Phi_j}\quad\quad\quad B^{k}_{ij}= \frac{\partial F^k_i}{\partial \Phi_j}.
\end{equation}
It is possible to find $U_i$ and $F^k_i$ if the matrices 
\begin{equation}
    \frac{\partial A_{ij}}{\partial \Phi_m},\quad\quad\quad \frac{\partial B^{k}_{ij}}{\partial \Phi_m}
\end{equation}
are symmetric with respect to the exchange of indices $j$ and $m$. 
If this was the case, one could rewrite the equations \eqref{eq:quasilinear} in the form \eqref{eq:conservativeForm}, and a solution with a piecewise constant initial condition could be defined. 

For Israel-Stewart evolution equations, the integrability condition is satisfied only for indices that correspond to the conservation of the energy-momentum tensor or baryon number current, but not for indices corresponding to the evolution equations for bulk viscous pressure, shear stress, or diffusion current. (This can be verified explicitly by examining the matrix provided in ref.~\cite{Floerchinger:2017cii}.) Unfortunately, a violation of the integrability conditions and associated conservation-type form makes the solution of the equation with a piecewise discontinuous initial condition ambiguous.

Even with an entropy condition supplemented to the equation, one needs to specify how to define the non-linear product between the matrix $B^{i}$ and the derivative across discontinuous data $\partial_k\Phi_j$ \cite{Maso1995DefinitionAW}. It is therefore possible that the current setup for dissipative relativistic fluids needs to be reconsidered if it is indeed necessary to work with discontinuous weak solutions.

Let us add here some remarks on the limit of very large relaxation times $\tau_\text{shear}, \tau_\text{bulk}, \tau_\text{heat} \to \infty$. This corresponds formally to infinite mean free time. With the initial condition of vanishing stress fields, $\pi_\text{bulk} = \nu^\rho = \pi^{\alpha\beta}=0$, the Israel-Stewart eqs. \eqref{eq:IS}, \eqref{eq:IS2} and \eqref{eq:IS3} have then a simple formal solution: the stress fields vanish at all times. This implies vanishing entropy production and the evolution equations for the energy density $\epsilon$, baryon density $n$ and the fluid velocity $v$ are then like for an ideal fluid. For vanishing transverse gradients this leads to a solution that has not only the longitudinal reflection symmetry $z\to -z$, but that is also symmetric with respect to time reflections $t\to -t$, where time is measured relative to the point of full overlap. At late times we find accordingly a situation that is analogous to the interactionless collision model investigated in section \ref{sec:InteractionlessColl}, which showed full transparency.

There are two modifications that become important for a more realistic description. First, when the relaxation times $\tau_\text{shear}$, $\tau_\text{bulk}$, and $\tau_\text{heat}$ remain finite, the stress fields $\pi_\text{bulk}$, $\nu^\rho$ and $\pi^{\mu\nu}$ adopt non-vanishing values and entropy is being produced. Second, transverse gradients become important, and lead to the usual expansion and dilution in the transverse plane.

\subsection{Isotropic collision model}

An essential feature of the early longitudinal collision dynamics is a strong compression followed by a subsequent expansion. During both stages, entropy is produced by bulk viscous dissipation. While this compression and expansion are supplemented by shear flow, it might be instructive to study a simplified model of isotropic compression and expansion where bulk viscous dissipation is dominating. This is the purpose of the present subsection. 

We consider a simple model, namely a universe homogeneously filled with nuclear matter, with its spacetime metric reading
\begin{align}
    \ud s^2 = - \ud t^2 + a(t)^2 \left( \ud x^2 +\ud y^2 +\ud z^2 \right),
\end{align}
where $a(t)$ is the time-dependent scale factor\footnote{Notice that this metric can only be mapped to a flat Minkowski space-time for certain $a(t)$ and is therefore in general not equivalent to a contracting/expanding sphere embedded in three-dimensional Minkowski space-time.}. Assuming invariance with respect to spatial translations and rotations, the fluid fields depend only on time, $\Phi(t,x,y,z)=\Phi(t)$. The invariance under spatial rotations ensures furthermore that vector fields can only have non-zero time components. The fluid velocity is then trivial, $u^\mu = (1,0,0,0)$, and the diffusion current vanishes due to its orthogonality to the fluid velocity. Similarly, the shear-stress tensor vanishes. The remaining non-vanishing fluid fields are the temperature, chemical potential, and bulk viscous pressure, $\Phi=(T,\mu, \pib)$. The equations of motion in this setup read,
\begin{align}
    \partial_t \epsilon + 3 H ( \epsilon +p+\pib) &=0 \nonumber,\\
    \partial_t n + 3H n &=0 ,\label{eq:PiHubble} \\
    \tau_\mathrm{bulk} \partial_t \pib + 3 H \zeta \nonumber\\
    + \left( 1 + 3H\frac{\tau_\mathrm{bulk}}{2} -\frac{\tau_\mathrm{bulk}}{2} \frac{\partial_t T}{T} \right) \pib  &=0. \nonumber
\end{align}
These equations are supplemented by Eq.~(\ref{eq:EvolutionVolumeRatioParam}) and Eq.~(\ref{eq:ModifiedEvolutionCoords}) in the presence of a first-order phase transition. 

We will model the time-dependent scale factor $a(t)$ such that the Hubble rate,
\begin{align}
    H(t) = \frac{\dot a(t)}{a(t)} = -\frac{\partial_t n}{3n},
\end{align}
corresponds to the time-dependent baryon density in the model of interactionless collisions in the center of the fireball, as investigated in section Sec.~\ref{sec:InteractionlessColl}, see Fig.~\ref{fig:LongitudinalLandauMatching}. With this choice, we obtain a simplified model for the main effect of fluid compression at early times. 

\begin{figure}[h]
\includegraphics[width=0.38\textwidth]{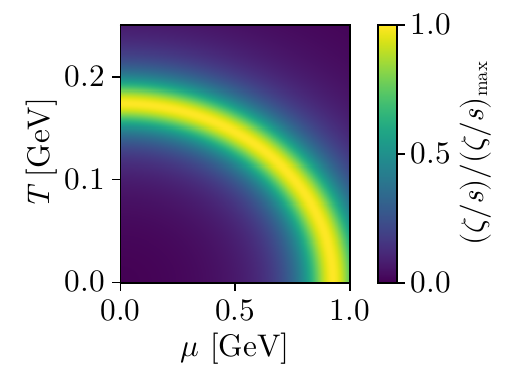}
\caption{Ratio of bulk viscosity and its amplitude. A standard parametrization for the bulk viscosity employed in calculations for $\mu=0$ has been extended for finite chemical potentials, such that the maximum of the viscosity coincides with the phase transition at $T=0$.}
\label{fig:Zetaparam}
\end{figure}

To fix the bulk viscosity we employ a standard parametrization \cite{Moreland:2014oya} and extend it to non-vanishing chemical potentials by substituting $T \to \sqrt{T^2+k ^2\mu^2}$. Here the parameter $k$ is chosen such that the peak of the bulk viscosity lies on the phase transition for $T=0$. The bulk viscosity and relaxation times are then given by
\begin{align}
    \zeta/s & = \frac{(\zeta/s)_\text{max}}{1+ \left( \frac{\sqrt{T^2+k ^2\mu^2}- \SI{24}{\MeV}}{\SI{175}{\MeV}} \right)^2}, \\
    \frac{\zeta}{\tau_\mathrm{bulk}(\epsilon+p)} &= \frac{\beta_0}{2},
\end{align}
where $\beta_0$ is a constant. The ratio $(\zeta/s)/(\zeta/s)_\text{max}$ is shown in Fig.~\ref{fig:Zetaparam} as a function of temperature and chemical potential.
Notice that the ratio $\zeta/\tau_\mathrm{bulk}$ vanishes at $T=\mu_{\rm crit}=0$, since both $\epsilon$ and $p$ vanish, ensuring the applicability of the approach as discussed in Sec.~\ref{sec:Hydrodynamics}.

We now dispose of all the necessary ingredients to solve the differential equations of motion for the fluid fields. The solutions for varying collision energies and viscosities are discussed in the next section.

\section{Results} \label{sec:Results}

\begin{figure}[h]
\includegraphics[]{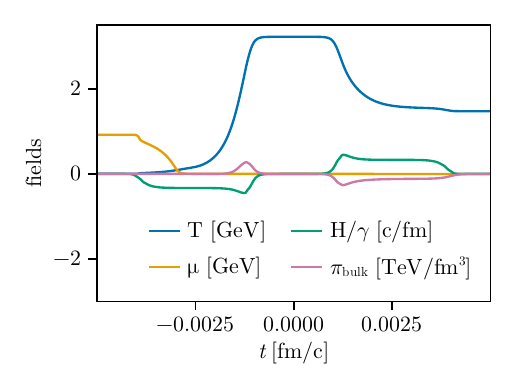}
\caption{Solution for the fluid fields as a function of time under a contraction and expansion phase. During the contraction phase, the baryon asymmetry is balanced. After that, a fast heating-up is observed due to internal friction. During the subsequent expansion, the system becomes more dilute and cools down. At late times the temperature remains constant at a finite value, which is larger than its initial value because of dissipation.}
\label{fig:OneEventSolution}
\end{figure}

We consider a heavy-ion collision at a center-of-mass energy per nucleon pair of $\sqrt{s_{\rm NN}}=\SI{2.76}{\TeV}$.
All fields except the chemical potential are initialized to zero. The chemical potential is initialized on the liquid-gas phase transition of the nucleon-meson model, $\mu (t=t_0)=\mu_\text{crit}=\SI{920}{\MeV}$, with $t_0=\SI{-0.006}{\femto\meter /c}$. A discussion about the initialization of the volume ratio parameter can be found below. The solutions are obtained by solving the system of ordinary differential equations in Eq.~\eqref{eq:PiHubble} with a Rosenbrock method in the Julia coding language \cite{bezanson2017julia,Bhagavan2024,pal2024nonlinearsolve,DifferentialEquations.jl-2017,RevelsLubinPapamarkou2016,DanischKrumbiegel2021}. 
The solution for the fluid fields, together with the Hubble rate, is shown in Fig.~\ref{fig:OneEventSolution}. We again define $t=0$ as the moment of maximal density (``full overlap"). During the initial contraction phase, the baryon asymmetry is balanced out, and the system transitions from $\mu=\mu_{\rm crit}$ to $\mu\approx0$. Before $\mu \approx 0$ is reached, the temperature only increases slowly. However, with the continuation of the contraction phase, the bulk pressure increases to counteract the externally forced contraction, resulting in a quicker heating-up. With the Hubble rate approaching zero again at the turning point at $t=0$, the bulk viscous pressure also relaxes to zero, with a delay compared to the Hubble rate given by $\tau_\mathrm{bulk}$. At the turning point, i.e. $H=0$, the fluid fields remain constant. In the subsequent expansion phase, the nuclear matter is already at $\mu \approx 0$, resulting in an immediate reaction of the bulk viscous pressure, which becomes negative in response to the expansion. The cooling goes along with the expansion, albeit not reaching $T=0$ due to the non-vanishing bulk viscosity\footnote{Notice here the difference with the case of the absence of interactions, where the evolution was symmetric under time-reversal symmetry by construction.}. Finally, when the expansion phase comes to an end, the chemical potential and temperature remain constant with $T(t=t_0)<T(t=t_f)$, while the bulk viscous pressure relaxes to its equilibrium value.

\begin{figure}[h]
\includegraphics[]{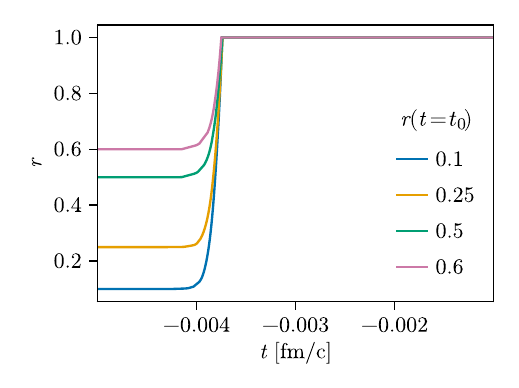}
\caption{Evolution of the volume ratio parameter for different initialization values. The rapid expansion of the system leads to it completing the phase transition quickly. The ensuing dynamics move the system away from the phase transition, allowing again for a complete thermodynamic description in terms of $T$ and $\mu$.}
\label{fig:OrderParamPlot}
\end{figure}

\begin{figure}[h]
\includegraphics[]{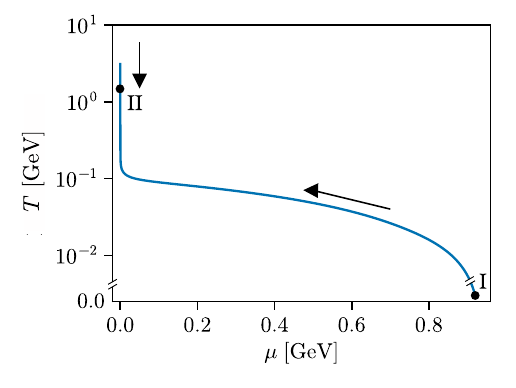}
\caption{Trajectory of a contracted and subsequently expanded homogeneous fluid through the phase diagram. The starting point of the evolution lies on the phase transition line of the nucleon-meson model (point I). During the initial contraction, the baryon asymmetry is balanced out first, until $\mu\approx 0$ is reached with a faster subsequent heating-up. The ensuing expansion phase, leading to a decrease in temperature, ends in the final point of the evolution (point II) at $\mu\approx 0$ and $T \approx \SI{750}{\MeV}$.}
\label{fig:basePhaseDiagPlot}
\end{figure}

Now we briefly address the initialization of the volume ratio parameter $r$. Its time evolution for different initial values is shown in Fig.~\ref{fig:OrderParamPlot}. Independently on the initial value, the rapid initial compression results in the volume ratio quickly approaching one, with all fluid in the dense phase, and then remaining constant for the rest of the evolution, because the matter heats up and moves away from the phase transition. Therefore, we can neglect the volume ratio parameter in the following discussions.
The trajectory of the fluid through the phase diagram is shown in Fig.~\ref{fig:basePhaseDiagPlot}. The starting point of the evolution lies on the phase transition line of the nucleon-meson model (point I). During the initial contraction, the baryon asymmetry is balanced out first, until $\mu\approx 0$ is reached with a faster subsequent heating-up. The ensuing expansion phase, leading to a decrease in temperature, ends in the final point of the evolution (point II) at $\mu\approx 0$ and $T \approx \SI{750}{\MeV}$. This corresponds to the peak temperature expected in an ultra-central Pb-Pb collision at $\sqrt{s}=\SI{2.76}{\TeV}$.

\begin{figure}[h]
\includegraphics[]{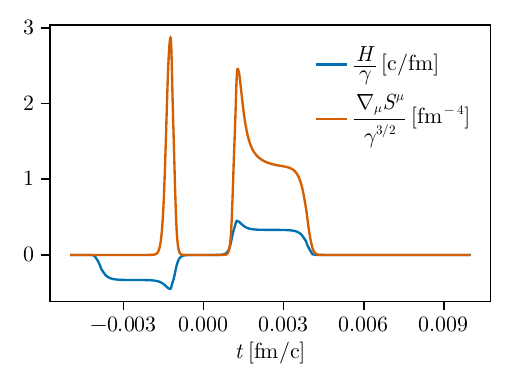}
\caption{Entropy production together with the Hubble rate as a function of the evolution time. The entropy production begins after the initial baryon asymmetry is balanced out, accounting for the delay of the entropy production compared to the Hubble rate. Afterward, the entropy production follows the Hubble rate.}
\label{fig:entropyProductionOneEvent}
\end{figure}

\begin{figure}[h]
\includegraphics[]{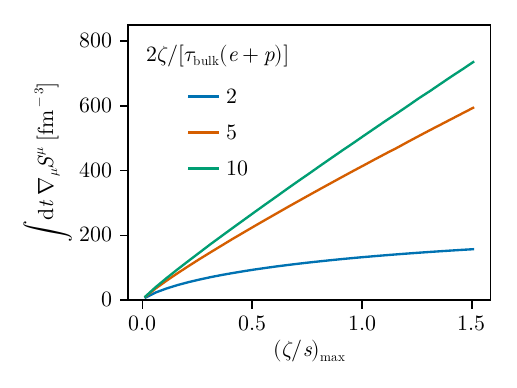}
\caption{Total produced entropy as a function of the bulk viscosity for different relaxation times. A larger viscosity leads to more entropy being produced due to the increased internal friction. Shorter relaxation times allow a faster reaction to changes in the scale factor, resulting in more entropy being produced.}
\label{fig:entropyProductionTotalZeta}
\end{figure}

\begin{figure}[h]
\includegraphics[]{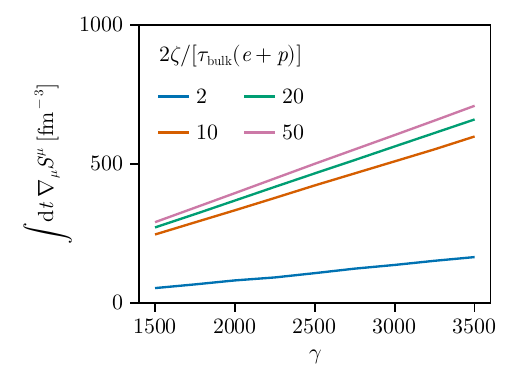}
\caption{Total produced entropy as a function of the Lorentz-$\gamma$ factor for different relaxation times. Collisions at higher energy (i.e. larger $\gamma$), more entropy gets produced. For large enough relaxation times, the entropy production loses efficiency by a slower reaction to the external work being done.}
\label{fig:entropyProductionTotalGamma}
\end{figure}

\begin{figure}[h]
\includegraphics[scale=1.0]{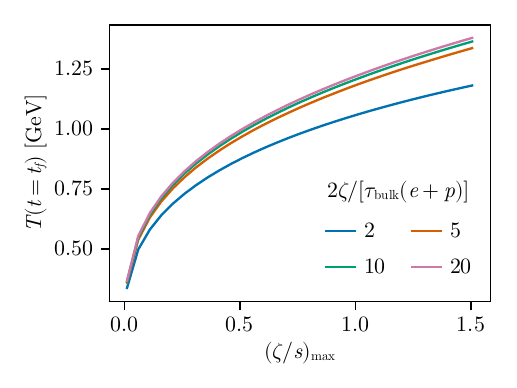}
\caption{Final temperature after the evolution as a function of the bulk viscosity for different relaxation times. Larger viscosities and shorter relaxation times result in effectively more viscous matter, in which the final temperature is larger than the initial temperature due to its internal friction.}
\label{fig:finalTemperatureZeta}
\end{figure}

Note that the fact that the nuclear matter moves from point I to point II during a compression-expansion cycle is a result of bulk viscous dissipation and the associated entropy production. An ideal fluid would not follow such a trajectory in the phase diagram because entropy would be conserved. At the same time, Navier-Stokes' equations would not allow for a causal and stable evolution of the fluid fields. The role of Israel-Stewart-like equations is therefore crucial to successfully describe a heavy-ion collision in its entirety through fluid dynamics.
The entropy production, computed as given in Eq.~(\ref{eq:entropy}), as a function of the evolution time is shown in Fig.~\ref{fig:entropyProductionOneEvent} together with the Hubble rate $H$. The entropy production is delayed with respect to the beginning of the contraction phase due to the initially present baryon asymmetry of nuclear matter. As soon as this asymmetry has been largely balanced out, the entropy production follows the Hubble rate.
Since the entropy in the initial state is strongly correlated with the number of particles produced in the final state (see e.g. \cite{Giacalone:2020ymy}), it is compelling to study the entropy production as a function of the bulk viscosity and its relaxation time, as well as its dependence on the collision energy. The total produced entropy (calculated as time integral of the entropy production rate) as a function of the bulk viscosity for different relaxation times (given by different values of $\beta_0$) is shown in Fig.~\ref{fig:entropyProductionTotalZeta}. In the limit of ideal fluid dynamics (i.e. vanishing viscosity), no entropy is produced. The larger the viscosity, the more entropy is produced. Furthermore, a shorter relaxation time (i.e. a larger $\beta_0$) enables a faster reaction to the external work being done on it by building up more bulk pressure. This results, in turn, in producing more entropy.
The relation between the relaxation time and the time in which the system expands/contracts is significant for the entropy production, as can be seen when studying the entropy production as a function of the collision energy, or equivalently, its $\gamma$ factor (Fig.~\ref{fig:entropyProductionTotalGamma}). The produced entropy scales with the Lorentz-gamma factor. This is expected from experimental measurements of charged hadrons' multiplicities, where a higher number of particles is produced at larger collision energies. Note that for the largest relaxation time, significantly less entropy is being produced. In this case, the entropy production loses efficiency, because the bulk viscous pressure does not fully relax to zero when the Hubble rate vanishes after the initial contraction phase. Therefore, less negative bulk pressure is created during the expansion phase, in turn leading to less entropy being produced.
As a final step, we want to examine the endpoint of the trajectory in the phase diagram (point II in Fig.~\ref{fig:basePhaseDiagPlot}). Since the evolution always ends up at $\mu \approx 0$, the main point of interest is the final temperature $T(t=t_f)$, with $t_f=\SI{0.006}{\femto\meter /c}$. The latter is shown as a function of $(\zeta/s)_\text{max}$ in Fig.~\ref{fig:finalTemperatureZeta}. In the limit of an ideal fluid with vanishing viscosity, we recover that the initial and final temperatures are the same $T(t=t_0)=T(t=t_f)=0$. Similarly to the produced entropy, the final temperature increases with increasing viscosity and increasing $\beta_0$, i.e. decreasing relaxation time, as a result of a fast and effective reaction to the external work (contraction and expansion). We find that if the relaxation time is sufficiently small, the final temperature seems to be almost independent of its specific value. In contrast to the produced entropy, we find a much steeper decrease in the final temperature when approaching the limit of vanishing viscosity, whereas the produced entropy shows an almost linear scaling with the viscosity for small enough relaxation times. This different scaling is consistent with the scaling of the entropy density as a function of the temperature, $s \propto T^3$. Furthermore, we want to stress that the scaling of the final temperature allows us to choose a combination of a viscosity together with a relaxation time, such that $T(t=t_f)=T_f$ for any value of $T_f$. Therefore, it is possible to match $T_f$ to temperatures of the initial profile found in other studies. Albeit not showing much constraining power in the current setting due to the lack of shear viscous dissipation and heat conductivity, this matching could offer interesting possibilities for constraining the transport coefficients, when applied in a more realistic scenario.

\section{Conclusions and outlook} \label{sec:Outlook}

We have investigated here the possibility of describing the soft part of an entire high-energy nuclear collision, starting from an initial state before the collision, in terms of relativistic fluid dynamics.
In particular, we have argued that the incoming nuclei in the vacuum that surrounds them can be seen, to a reasonable approximation, as droplets of cold nuclear matter at the first-order nuclear liquid-gas transition. To better understand the actual collision dynamics itself we have employed first a simple model where the energy-momentum tensor and baryon number currents of the two freely propagating nuclei are linearly superposed. This corresponds to a hypothetical limit with vanishing cross sections. The combined currents were then decomposed, using Landau frame matching, to determine energy and baryon number densities, the fluid velocity, shear stress, bulk viscous pressure, and diffusion current fields as functions of space and time. Besidews the longitudinal reflection symmetry $z \to -z$, this model also shows a time reflection symmetry, $t \to -t$, where time is measured relative to the point of full overlap, characterizing the full transparent limit. Within this model, one finds very sharp features for the fluid fields as a function of longitudinal coordinates and time, suggesting that shock-like phenomena are to be expected.

As a preparatory step to discuss fluid evolution equations, we have discussed properties of the thermodynamic equation of state and interpolation between different analytically or numerically known expressions. Important for our purpose is a realistic description both in the vicinity of the nuclear liquid-gas phase transition and at high temperature and small baryon chemical potential, as well as a reasonably accurate description for the region in between. At large temperatures and small baryon chemical potential, we are using numerical results from lattice QCD simulations employing a Taylor expansion around vanishing chemical potential. For smaller temperatures, as well as for larger values of the baryon chemical potential, we are employing a hadron resonance gas approximation. Finally, in the vicinity of the nuclear liquid-gas transition we are using a phenomenological nucleon-meson model which we solve in mean-field approximation. All these models have been combined in terms of interpolations.

We have then discussed fluid evolution equations that can potentially be used throughout the phase diagram. The conservation laws for energy and baryon number directly lead to evolution equations for the energy density and baryon number density or the conjugate thermodynamic variables temperature and chemical potential. We have here extended the formalism such as to allow for the evolution through first-order phase transitions. Without going into the more sophisticated physics of the phase transition we have employed here a relatively simple picture, that is in principle only suitable for slow enough dynamics and large length scales. In this picture the phase transition is assumed to happen instantaneously at the phase coexistence line and it suffices to introduce one additional parameter (the volume ratio of fluid in the two phases) to describe the fluid evolution during the transition.

In order to complete the equations of motion we have revisited relativistic fluid dynamics at second order in gradients, in the formulation first proposed by Israel and Stewart. This formulation features an entropy current and a local form of the second law of thermodynamics with non-negative local production of entropy. We believe that this form of local entropy production by shear and bulk viscous dissipation, as well as heat conduction, is in principle capable to account for the entire entropy that is necessarily produced during a heavy-ion collision event.

We have argued that for the very first moments of the actual collision event one can reduce the dynamical evolution equations to the longitudinal direction along the beam axis, as well as time, because gradients in the transverse direction are much smaller and become important only slightly later. The resulting equations of motion are quasi-linear hyperbolic evolution equations for six independent fluid fields.

While these evolution equations, resulting from the Israel-Stewart theory, can in principle be initialized also far from equilibrium, some theoretical uncertainty arises if shock-like phenomena should develop. In such a situation, where classical solutions to the equations of motion loose there meaning and one must look for weak solutions, it would be preferable to have a divergence-type formulation of the equations of motion at hand. At present we have to leave this point open and further investigations in future work will have to settle whether weak solutions are needed here, indeed.

An interesting limit within the Israel-Stewart description of the early longitudinal dynamics occurs at infinite relaxation times corresponding formally to diverging mean free time. The bulk viscous pressure, shear stress and diffusion current vanish then not only at early but at all times. In that situation there is no entropy production, and the evolution equations for the energy density, baryon density and longitudinal fluid velocity are the same as for an ideal fluid. For vanishing transverse gradients this solution is time-reflection symmetric, and resembles therefore at late times the interactionless collision limit with full transparency.

In order to study entropy production in a somewhat simpler setup we have also developed a reduced model for the strong and very quick compression and expansion of matter happening at the center of colliding nuclei. This model features an isotropic contraction and expansion for homogeneous matter and can be cast in the form of evolution equations in a hypothetical Friedmann-Lemaitre-Robertson-Walker universe. We have adapted the time-dependent scale factor such that it corresponds to the center of a central collision in the interactionless collision model described earlier. In this setup one can study indeed the evolution throughout the full phase diagram and quantify the production of entropy by bulk viscous dissipation. 

In this setup, we were able to study the increase of bulk viscous pressure and concurrent entropy production. We find that initially the entropy production lags behind the external compression due to the initial baryon asymmetry. After this has been balanced out, the bulk viscous pressure follows the external work with barely any delay. The production of entropy through dissipation results indeed in a large temperature and small chemical potential, as expected. We were able to show that it is possible to choose a set of parameters for the maximal bulk viscosity, and its relaxation time, such that the final temperature is close to the ones found in the center of the fireball created during a heavy-ion collision. Clearly, at this stage, this model is too simple to capture the real dynamics of a heavy-ion collision. However, our findings provide strong indications that a more realistic model can be developed to describe the initial stages of the soft part of a heavy-ion collision within relativistic viscous fluid dynamics.

A strong contender for such a more realistic model is given by the two nuclei with infinite extension in the transverse plane (equivalent to neglecting transverse gradients for the first instances of the collision) outlined in Sec.~\ref{sec:LongitudinaCollision}. Both the shear-stress tensor and baryon diffusion current can here be non-vanishing. Practically the transition to this longitudinal setup includes several challenges to overcome: Firstly, the evolution equations are no longer a set of ordinary differential equations, but rather a set of coupled partial differential equations that make the numerical treatment more involved. The large $\gamma$-factors together with the interplay of the viscous corrections and the high compression in the initial moments of the collision require especially careful treatment of the numerical implementation. An additional requirement for the longitudinal collision is the set of shear and baryon diffusion transport coefficients as functions of the temperature and chemical potential. So far, the calculation of these from first principles, especially at high chemical potentials and vanishing temperatures, is rather involved. Therefore, our work presents a motivation to improve this. 

Before comparing theoretical calculations of transport properties to fluid dynamic modeling of the initial state along the lines we propose here in future work, further investigations concerning 
the range of applicability of the theory and 
the practical application of the idea are needed. Another important step in the improvement of this model is the inclusion of fluctuations in the initial state, which are crucial in the explanation of several phenomena observed in heavy-ion collisions. This can either be achieved by introducing fluctuations in the initial density profiles (Eq.~(\ref{eq:InitalDensityWS})) on an event-by-event basis or with the inclusion of evolution equations of correlation functions of the initial density, such as the two-point function. We aim at the implementation of this second option in a continuation of this work.

\section*{Acknowledgements}
The authors want to thank A. Erschfeld for his contributions in the initial stages of the work and T. Stötzel and A. Sorensen for useful discussions. This work is part of and supported by the DFG Collaborative Research Centre SFB 1225 (ISOQUANT). A.K. is supported by the U.S. Department of Energy, Office of
Science, Office of Nuclear Physics, grant No. DE-FG02-05ER41367.

\appendix
\section{Thermodynamic equation of state} \label{sec:appendixEoS}

\subsection{Lattice QCD} \label{sec:LQCD}

For the region of high temperature and small baryon chemical potentials, the equation of state can be calculated using the methods of lattice QCD \cite{Borsanyi:2016ksw,Borsanyi:2022qlh,HotQCD:2014kol}. We use an interpolation given by an equilibrium pressure at vanishing baryon chemical potential together with an expansion to sixth order in chemical potential

\begin{equation}
    p(T,\mu) = p(T) + \sum_{n=2,4,6} \frac{\chi_n(T)}{n!} \mu^n.
\end{equation}

Since the interpolations given in the literature are obtained using the Padé approximation, they become unphysical due to their pole at low temperatures. This pole can be removed by switching to the hadron resonance gas, describing QCD matter at low temperatures and densities.

\subsection{Hadron resonance gas} \label{sec:HRG}

The pressure of the hadron resonance gas \cite{Huovinen:2009yb} is given by the sum of all the partial pressures of the appearing resonances

\begin{align}
\PHRG(T, \mu) = &\sum_{\textrm{baryons}} d_i \PFG(T, B_i \mu; m_i) \\
+&\sum_{\textrm{mesons}} d_i \PBG(T, 0; m_i) \; ,
\end{align}

where the $d_i$ are degeneracy factors related to spin, color charge, etc., and $B_i$ being the baryon charge. The partial pressures are given by

\begin{equation}
\PFG(T, \mu; m) = \frac{\left( mT \right)^2 }{2 \pi^2} \sum_{k=1}^\infty \frac{(-1)^{(k+1)}}{k^2} \, K_2\left( \frac{km}{T} \right) \, e^{\frac{k \mu}{T}} \; ,
\end{equation}

for fermions and 

\begin{equation}
\PBG(T; m) = \frac{(mT)^2}{2 \pi^2} \sum_{k=1}^\infty \frac{1}{k^2} \, K_2\left( \frac{km}{T} \right) \; 
\end{equation}

for bosons. Here we sum over all hadron resonances with mass $< \SI{2.1}{\GeV}$ \cite{Alba:2017hhe,Alba:2017mqu,Alba:2020jir}. Since the derivation of the hadron resonance gas is based on the virial expansion, it is only a valid description at low densities. Therefore, we need an additional model covering the low temperature, high-density area, including the first-order phase transition from vacuum to nuclear matter. A suitable model describing this range of the QCD phase diagram, including the phase transition is a nucleon-meson model.

\subsection{Nucleon-meson model} \label{sec:Walecka}

The nucleon-meson model~\cite{WALECKA1974491,Eser:2023oii} is an effective model for cold nuclear matter interacting via the exchange of scalar and vector mesons. In general, linear nucleon-meson models include the interactions of the scalar meson $\sigma$, the pseudo-scalar mesons $\pi^0,\pi^\pm$ and the singlet vector meson $\omega_\mu$ with baryonic matter $\psi_a$, where the index $a$ runs over protons and neutrons. However, for our purposes a simplified model is sufficient. Therefore, we will assume that the nuclear matter is isospin symmetric and neglect the pseudo-scalar interaction. The effective Lagrangian is then given by

\begin{align}
    \mathcal{L} = &\bar{\psi} i \gamma^\nu \left( \partial_\nu - i g \omega_\nu - i\mu \delta_{0\nu} \right) \psi + h \sigma \bar{\psi} \psi \nonumber\\
    &+ \frac{1}{4} \sigma \left( -\partial_\mu \partial^\mu  \right) \sigma+ U_\text{mic} (\sigma, \omega_0) \\
    &+ \frac{1}{4} F_{\mu\nu}F^{\mu\nu} + \frac{1}{2} m^2_\omega \omega_\mu \omega^\mu \nonumber
\end{align}

with the microscopic potential being parametrized as

\begin{align}
    U_\text{mic}= &\frac{1}{2} m_\pi^2 (\sigma^2 -f_\pi^2) + \frac{1}{8} \lambda (\sigma^2 -f_\pi^2)^2 \nonumber\\
    &+ \frac{1}{3} \frac{\gamma_3}{f_\pi^2} (\sigma^2 -f_\pi^2)^3 + \frac{1}{4} \frac{\gamma_4}{f_\pi^4} (\sigma^2 -f_\pi^2)^4 \\
    &- m_\pi^2 f_\pi (\sigma -f_\pi) -\frac{1}{2} m_\omega^2 \omega_0^2 \nonumber.
\end{align}

The appearing parameters are the couplings $g$ and $h$ to the nuclear matter, the chemical potential $\mu$, the mass of the omega and sigma mesons $m_\omega$ and $m_\sigma$ together with the pion decay constant $f_\pi$ and the other potential parameters $\lambda$, $\gamma_3$, and $\gamma_4$.

We fix these parameters by using the experimentally measured values for the omega mass $m_\omega = \SI{783}{\MeV}$, the pion mass $m_\pi = \SI{135}{\MeV}$ and the pion decay constant $f_\pi  = \SI{93}{\MeV}$. The coupling constant $h$ can be fixed by the requirement $h f_\pi = m_N$, with $m_N=\SI{939}{\MeV}$ being the nucleon mass, yielding $h=10$. The remaining parameters can be fixed by the constraints given from the first-order phase transition together with constraints relating to the properties of nuclear matter. The values we will be using are given by $g=9.5$, $\lambda = 50$, $\gamma_3 = 3$ and $\gamma_4 = 50$.

Using the mean field approximation, the pressure of the nucleon-meson model then is given by the Fermi pressure of the nucleons modified by the presence of the sigma and omega mesons

\begin{align}
    \PWM = 4 \PFG (T, \mu^*, m_N^*) - U_\text{mic}(\Bar{\sigma},\Bar{\omega}_0)
\end{align}

where $\Bar{\sigma}$ and $\Bar{\omega}_0$ are the mean field values of the meson fields. Note that the chemical potential and the nucleon mass entering the Fermi pressure are modified by the presence of the sigma and omega fields. The effective chemical potential is given by $\mu^* = \mu + g \Bar{\omega}_0$. Similar to the matching condition used to fix the parameters, the effective nucleon mass is given by $m_N^* = h\bar{\sigma}$. The values of the meson fields $\Bar{\sigma}(T,\mu)$ and $\Bar{\omega}_0 (T,\mu)$ can be determined self consistently by the so-called gap equations

\begin{align}
    \partial_{\Bar{\sigma}} \PWM (T,\mu^*,m_N^*) &=0, \\
    \partial_{\Bar{\omega}_0} \PWM (T,\mu^*,m_N^*) &=0.
\end{align}

\begin{figure}[h]
\includegraphics[]{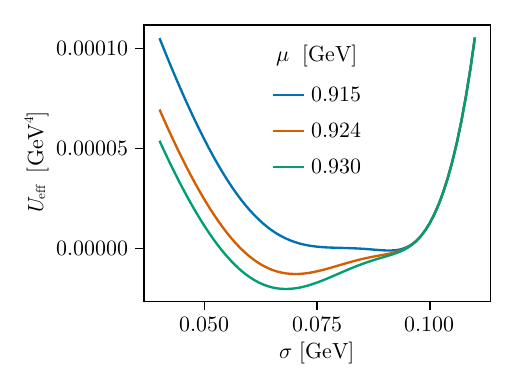}
\caption{Effective potential of the nucleon-meson model evaluated for different values of the $\sigma$-meson field. The field value for the $\omega_0$ field is obtained by solving the gap equations for a fixed value of $\sigma$. The effective potential shows two minima. For chemical potentials smaller than the critical chemical potential, the system is in the right minimum with $U_\text{eff}=-p=0$. At the critical chemical potential, the left minimum becomes degenerate with the previously global one on the right. For even larger values of the chemical potential, the left minimum becomes the global minimum with $U_\text{eff}=-p<0$. The discontinuous transition of the $\sigma$-field gives rise to the first-order phase transition.}
\label{fig:WaleckaPotential}
\end{figure}

The resulting baryon number density then has a jump for temperatures below $T_* \approx \SI{20}{\MeV}$. For temperature above $T_*$, the number density is a smooth function. The effective potential evaluated for fixed $\bar{\sigma}$ is shown in Fig.~\ref{fig:WaleckaPotential}. The effective potential shows two minima. For chemical potentials smaller than the critical chemical potential, the system is in the right minimum with $U_\text{eff}=-p=0$. At the critical chemical potential, the left minimum becomes degenerate with the previously global one on the right. For even larger values of the chemical potential, the left minimum becomes the global minimum with $U_\text{eff}=-p<0$. The discontinuous transition of the $\sigma$-field gives rise to the first-order phase transition.

\clearpage
\bibliography{references}
\clearpage
\end{document}